%% file: neurips_2026.tex
\documentclass{article}

 \usepackage[main, final]{neurips_2026}
\usepackage[utf8]{inputenc} % allow utf-8 input
\usepackage[T1]{fontenc}    % use 8-bit T1 fonts
\usepackage{hyperref}       % hyperlinks
\usepackage{url}            % simple URL typesetting
\usepackage{booktabs}       % professional-quality tables
\usepackage{amsfonts}       % blackboard math symbols
\usepackage{nicefrac}       % compact symbols for 1/2, etc.
\usepackage{microtype}      % microtypography
\usepackage{xcolor}         % colors

\hypersetup{hidelinks}
\usepackage{tipa} % for IPA
\usepackage{amsmath}
\usepackage{multirow}
\usepackage{enumitem}

\usepackage{colortbl}
\usepackage{wrapfig}
\usepackage{graphicx}
\usepackage{pifont}
\usepackage{pgf}
\usepackage{fontawesome5}
\usepackage{titletoc}

\definecolor{redx}{HTML}{A50E0E}
\newcommand{\redx}{\textcolor{redx}{\ding{55}}}

\usepackage{subcaption}

\title{Embedded Arena: Iterative Optimization via Hardware Feedback}

\newcommand{\ilogo}[1]{%
  \raisebox{2pt}{\includegraphics[height=5pt]{#1}}%
}

\newcommand{\UW}{\ilogo{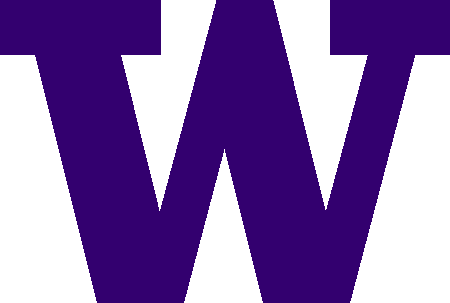}}
\newcommand{\UCSD}{\ilogo{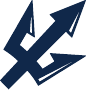}}
\newcommand{\NEU}{\ilogo{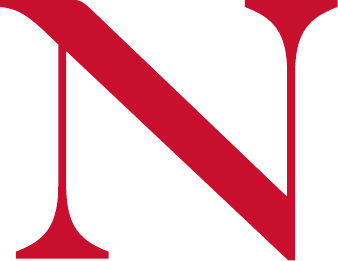}}

\author{
    \textbf{Zhihan Zhang}\UW$^{*}$ \quad
    \textbf{Alexander Le Metzger}\UW$^{*}$ \quad
    \textbf{Jiuyang Lyu}\UW$^{*}$ \quad
    \textbf{Chun-Cheng Chang}\UW \quad
    \vspace{0.2em} \\
    \textbf{Jiayi Shao}\UW \quad
    \textbf{Yujia Liu}\UCSD \quad
    \textbf{Emmanuel Azuh Mensah}\UW \quad
    \textbf{Edward Wang}\UCSD \quad
    \textbf{Kurtis Heimerl}\UW
    \vspace{0.2em} \\
    \textbf{Gregory D. Abowd}\NEU \quad
    \textbf{Shwetak Patel}\UW \quad
    \textbf{Natasha Jaques}\UW \quad
    \textbf{Vikram Iyer}\UW
    \vspace{0.5em} \\
    \UW University of Washington \quad \UCSD University of California San Diego \quad \NEU Northeastern University \\
    $^{*}$Equal contribution
    \vspace{0.2em} \\
    \textcolor[HTML]{A50E0E}{\textbf{NeurIPS 2026 Spotlight}}
}

\begin{document}

\maketitle

\begin{abstract}
As AI moves into the physical world, from robots to environmental sensors to medical wearables, it increasingly relies on local inference to meet latency, communication, or privacy constraints. Optimizing models for heterogeneous edge hardware such as microcontrollers (MCUs) requires simultaneously satisfying hard physical constraints on memory, power, and temperature while preserving accuracy, a multidimensional optimization that is today performed manually by experts. We ask whether an LLM agent can autonomously navigate this complex, multi-turn pipeline guided by real hardware feedback, and introduce a hardware-in-the-loop agent arena in which the agent iteratively refines both model and firmware by compiling, flashing, and measuring on real hardware, enabling closed-loop optimization. Frontier models, including Claude Opus 4.7 and Gemini 3.1 Pro, fail entirely without hardware feedback (0\% deployment success), whereas our hardware-in-the-loop formulation achieves the first successful deployment within three iterations and can surpass human expert results within seven. This agentic co-optimization achieves 250× compression for vision models with <3.3\% accuracy loss and 400× for audio with <6\% Feature Error Rate loss, enabling battery-free operation on a commercial MCU via solar harvesting. We demonstrate practical impact in two real-world systems: an elk-detection camera trap (96.7\% accuracy) and a phonetic-transcription wearable (8.37\% FER) for child development research.\footnote{\faGithub\hspace{0.5em}\href{https://ubicomplab.github.io/embedded-arena/}{\texttt{https://ubicomplab.github.io/embedded-arena/}}}
\end{abstract}

\section{Introduction}
Advances in large, cloud-based AI models have unlocked remarkable capabilities across vision, language, and multimodal reasoning, but leave a gap for applications with stringent latency, privacy, or connectivity requirements. Remote wildlife cameras must detect and classify species of interest without reliable network access~\cite{desai_camaroptera_2022, dong_promise_2026}, mobile devices may process audio and biosignals locally to avoid the privacy risks and power costs of cloud streaming~\cite{ahmed_mlung_2019,chatterjee_mlung_2019,larson_accurate_2011}, and physical AI systems with safety-critical perception-action cycles cannot tolerate round-trip latency to the cloud. Such applications demand small models that run directly on microcontroller (MCU) class hardware. These devices have many orders of magnitude lower clock speeds, memory, and storage than datacenter GPUs, requiring substantial model compression and optimization to utilize the growing hardware acceleration capabilities being built into MCUs~\cite{curtiss_facebit_2022}.

Deploying AI in this heterogeneous, rapidly evolving hardware landscape requires multidimensional, system-level optimization that satisfies hard physical constraints (e.g., memory footprint, power consumption, thermal limits) while maximizing model accuracy. 
On the hardware side, this landscape is a moving target: the STM32F7~\cite{lin_mcunet_2020} MCU series alone ships in 22 silicon variants, each with distinct memory configurations and AI accelerators.
On the model side, practitioners must navigate compression techniques (e.g., quantization, operator substitution), each of which interacts differently with the target chip's processing capabilities.
These optimization choices are coupled and competing. For example, for object detection on MCUs, reducing camera resolution saves activation memory and power draw but directly degrades accuracy~\cite{bompani_multi-resolution_2024}.
Today, navigating these trade-offs is performed manually by experts who hold rare, joint expertise in embedded systems and machine learning (ML). Skilled practitioners using deployment frameworks such as LiteRT~\cite{google_litert_nodate} spend weeks optimizing a single model on real hardware for a single deployment---a process that must be repeated from scratch for every new chip~\cite{lin_mcunet_2020}. 
In this paper, we ask: \emph{can AI agents optimize AI to run on real hardware by iterating directly on physical measurements?}

To answer this question, we construct a \textbf{hardware-in-the-loop (HIL)} environment in which the agent iteratively produces, compiles, flashes, and refines firmware subject to physical feedback. We observe that even state-of-the-art models like Claude Opus 4.7 fail entirely in this setting when restricted to software-only feedback. 
Closing the loop with physical measurements, such as power and temperature directly from the device, guides the agent toward firmware that actually works on the device. HIL raises deployment success rates from 0\% to 80\% on the hardest tasks.
We formalize this loop as a continuous optimization problem by reducing physical measurements to a programmatically scorable scalar objective, enabling iterative refinement. This transforms an opaque, expert-gated process into an accessible workflow, letting practitioners without joint hardware-AI expertise deploy state-of-the-art models on resource-constrained devices.

Our HIL formulation also addresses a gap in existing agentic systems and LLM benchmarks.
While LLM-based agents have demonstrated impressive autonomy in complex workflows~\cite{zhang_sustainability_2026,kim_capable_2026,karpathy_autoresearch_nodate,yang_swe-agent_2024,nam_mle-star_2026,aygun_ai_2026,gottweis_accelerating_2026,chi_frontier-eng_2026}, benchmarks for evaluating them on long-horizon, system-level optimization remain rare.
Most (e.g., SWE-bench~\cite{jimenez_swe-bench_2023}, TerminalBench~\cite{merrill_terminal-bench_2025}, HumanEval~\cite{chen_evaluating_2021}) focus entirely on the digital domain, evaluating agents on isolated, human-curated single-turn tasks rather than end-to-end system optimization.
Because these tasks are scored on binary correctness, they lack a mechanism for continuous self-improvement and saturate quickly as models advance, requiring periodic manual updates. 
A smaller family of Kaggle-style competitions~\cite{aygun_ai_2026} poses open-ended optimization scored by continuous software metrics (e.g., held-out accuracy) to support iterative hill-climbing.
Our HIL arena inherits this open-ended optimization paradigm but extends it to performance metrics measured directly from the physical world at test time, where results vary with ambient conditions and component state, forcing agents to reason about real hardware behavior.

\begin{figure}[t]
    \centering
    \includegraphics[width=\linewidth]{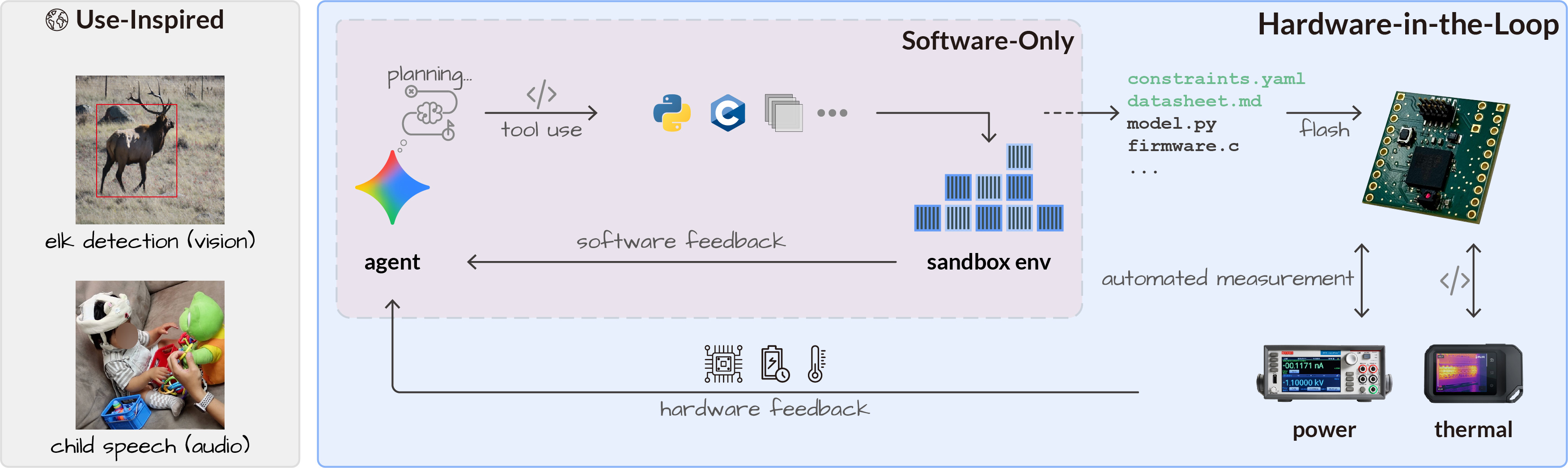}
    \caption{AI agents iteratively optimize models to run on MCUs within their tight power, thermal, and compute constraints using hardware-in-the-loop feedback. We demonstrate successful model optimization for real-world applications, including wildlife detection and child speech monitoring.} 
    \vspace{-1em}
    \label{fig:teaser}
\end{figure}

We summarize our contributions as follows:
\begin{enumerate}[leftmargin=*]
\item We are the first to formalize the deployment of AI on resource-constrained devices as a hardware-in-the-loop agent arena for iterative optimization, and physically build the end-to-end evaluation infrastructure.
\item We perform a comprehensive study of how context and hardware feedback shape the artifacts produced by LLM agents. 
HIL significantly improves deployment success over both software-only baselines, achieves the first successful deployment within three iterations, and \textbf{can surpass human expert results within seven}.
% Our analysis uncovers systematic failure modes, including XX, and yields concrete guidelines for developing more reliable LLMs for embedded systems development.
\item We achieve 250× compression for vision models with <3.3\% accuracy loss, and 400× for audio with <6\% Feature Error Rate (FER), also reducing power draw by over 58\% and peak device temperature rise by over 40\%. 
\item We demonstrate practical impact in two real-world deployments developed in direct collaboration with domain practitioners: a solar-powered camera trap for elk detection achieving 96.7\% accuracy under 7 mW average power, and a children's head-worn device with 8.37\% on-device FER for phonetic transcription. 
\end{enumerate}

% Taken together, our results support a broader vision: \emph{if a real-world physical objective can be reduced to a programmatic score, agents can optimize it iteratively.}

\section{Hardware-in-the-Loop Arena}\label{sec:hil_arena}
We present a hardware-in-the-loop arena in which an LLM agent must generate, compile, flash, and iteratively refine software subject to \emph{physical} feedback signals returned by real hardware.
Unlike software-only benchmarks, our arena closes the loop with the physical world: rewards are not computed by scoring functions, but are \emph{measured} by instruments attached to the device under test.
We now formalize this as a constrained optimization problem over real hardware, then instantiate it across three physically distinct axes: model compression, power minimization, and thermal management.
The formalism we develop in \S\ref{subsec:problem_form} is general to any constrained hardware target.

\setlength{\columnsep}{1.5em}
\begin{wrapfigure}{r}{0.3\textwidth}
    \centering
    \vspace{-2em}
    \includegraphics[width=0.3\textwidth]{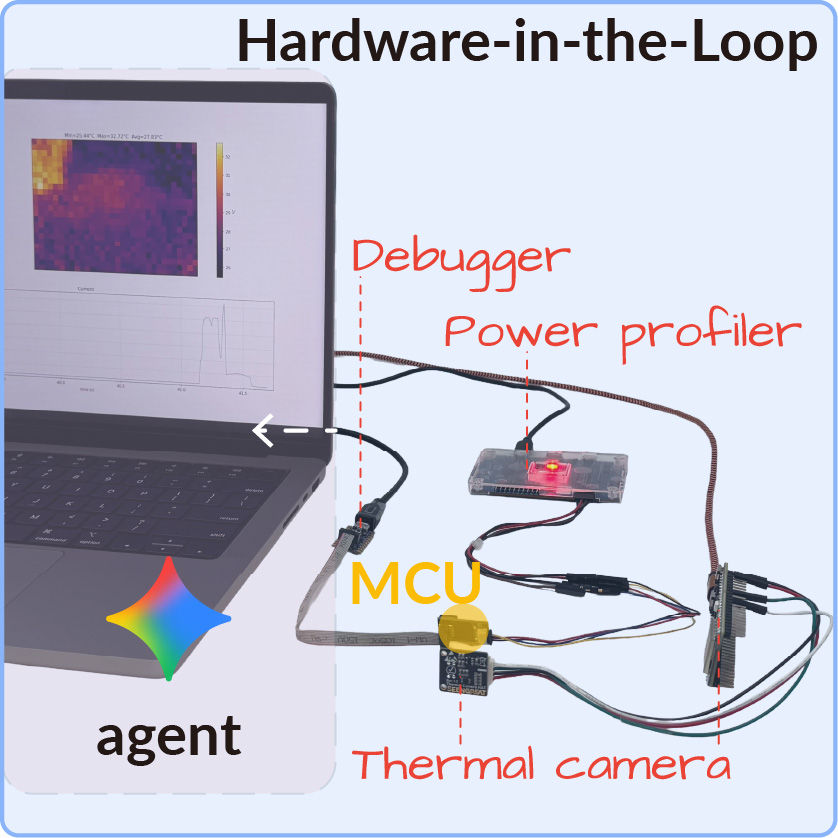}
    \caption{Hardware setup.} 
    \label{fig:hardware_setup}
    \vspace{-2em}
\end{wrapfigure}

\subsection{Problem Formalization}\label{subsec:problem_form}

\paragraph{Hardware constraint vector.}
Let $\mathcal{H}$ denote a target hardware platform characterized by a \emph{constraint vector}
\begin{equation}
    \mathbf{h} = (h_1, h_2, \ldots, h_K) \in \mathbb{R}^K,
    \label{eq:constraint-vec}
\end{equation}
where each component $h_k$ encodes a physical resource limit (e.g., flash storage in bytes, static RAM in bytes, peak supply current in amperes, maximum junction temperature in kelvin, supported operators).  
A solution is \emph{deployable} only if all $K$ constraints are simultaneously satisfied. 
We note that $\mathbf{h}$ can be automatically identified from datasheets via web search and verified using a combination of the compiler toolchain.

\paragraph{Solution space.}
A central design choice of our arena is that the solution space $\mathcal{S}$ is entirely open-ended. The agent produces a software artifact $s \in \mathcal{S}$, which may be firmware source code, a serialized neural network, a linker configuration, or any combination thereof.
The arena is \emph{agent-agnostic} and imposes no constraints on how $s$ is constructed: the agent may employ any architecture or strategy (e.g., retrieval-augmented generation, multi-agent architectures~\cite{kim_capable_2026}) and is judged solely on deployability and physical performance on the target hardware.

Given $s$ and $\mathbf{h}$, the build-and-flash operator $\Phi : \mathcal{S} \times \mathbb{R}^K \to \{0,1\}$ returns $1$ if $s$ compiles, boots successfully, and executes the required workload correctly on the device, and $0$ otherwise.
When $\Phi(s,\mathbf{h})=1$, a physical measurement $\mathcal{M} : \mathcal{S} \times \mathbb{R}^K \to \mathbb{R}^M$ returns a vector of $M$ empirically observed quantities $\mathbf{m} = \mathcal{M}(s, \mathbf{h})$ (e.g., inference accuracy, peak current draw, peak core temperature).

\paragraph{Optimization objective.}
Each benchmark task is defined by a tuple $\mathcal{T} = (\mathbf{h},\, f,\, \mathbf{c})$, where $f : \mathbb{R}^M \to \mathbb{R}$ is a task-specific objective to optimize, and $\mathbf{c} \in \mathbb{R}^C$ denotes task-level feasibility constraints that complement the hard physical limits in $\mathbf{h}$ and must be satisfied for $s$ to be valid (e.g., a minimum accuracy floor).
The agent's goal is to find a deployable artifact $s^{\star}$ that minimizes the task objective:
\begin{equation}
    s^{\star} = \operatorname*{arg\,min}_{s \in \mathcal{S}}
        \; f\!\left(\mathcal{M}(s, \mathbf{h})\right)
    \quad \text{s.t.} \quad
        \Phi(s, \mathbf{h}) = 1.
    \label{eq:opt}
\end{equation}
Because each exploration round consumes real resources (e.g., LLM inference, compilation, physical hardware measurements), the number of iterations $n$ before reaching a solution is penalized as a separate metric.

\paragraph{Iterative agent loop.}
The benchmark is structured as a multi-turn interaction
$\{s_0\} \cup \{(s_t,\, o_t)\}_{t=1}^{N_{\max}}$ (Figure~\ref{fig:teaser}).
At $t=0$, the agent is provided an initial artifact $s_0$ (e.g., a \texttt{model.py} specifying a pre-trained model architecture, or a \texttt{firmware.c} implementing a reference inference pipeline) together with the hardware constraint vector $\mathbf{h}$ and the task objective $f$.
The initial artifact $s_0$ serves as a starting point for optimization and need not satisfy the hardware constraints outright (e.g., $\Phi(s_0, \mathbf{h}) = 0$ when a pre-trained model exceeds the target flash budget).

At each subsequent step $t \geq 1$, the agent submits a modified artifact $s_t$; the benchmark executes $\Phi(s_t, \mathbf{h})$ and, if $\Phi=1$, runs $\mathcal{M}$, then returns the full observation $o_t = (\Phi(s_t,\mathbf{h}),\, \mathbf{m}_t)$ to the agent as a structured natural-language summary (e.g., current draw, measured chip temperature).
If $\Phi(s_t, \mathbf{h}) = 0$, the agent receives only the failure diagnostic without physical measurements.

The agent is allotted a budget of $N_{\max}$ attempts.

\paragraph{Agent memory and context.}
At each round $t$, the agent's artifact $s_t$ is written to a dedicated file (e.g., \texttt{iteration\_t.c}), preserving the full artifact history on disk. 
The observation $o_t$ is appended as a structured natural-language summary directly to the agent's context window (see Appendix \ref{sec:example_feedback_observation}). 
The agent is additionally equipped with tools to read and edit local files.
The agent may query all prior artifacts $\{s_0, \ldots, s_{t-1}\}$ on demand to inform $s_t$.
We preserve the full failure trajectory rather than maintaining only the current best artifact via version-control rollback~\cite{karpathy_autoresearch_nodate}, so the agent has the \emph{option} to understand past failures when proposing $s_{t}$.

\paragraph{Benchmark score.}
To enable fair comparison across agents, we define the benchmark score $\mathcal{B}$ for a task as
\begin{equation}
    \mathcal{B} = f\!\left(\mathcal{M}(s^{\star}, \mathbf{h})\right),
    \label{eq:benchmark-score}
\end{equation}
where $s^{\star}$ is the best deployable solution found by the agent within $N_{\max}$ rounds, and $f$ is defined so that lower $\mathcal{B}$ is always better (maximization objectives are negated). 
If no deployable solution satisfying $\mathbf{c}$ is found within $N_{\max}$ rounds, the agent receives $\mathcal{B} = \infty$.

When two agents achieve equal $\mathcal{B}$, we break ties by the total number of attempts $n \leq N_{\max}$, including attempts where $\Phi(s_t, \mathbf{h}) = 0$ (i.e., the artifact fails to compile or boot). Lower $n$ is better.

\subsection{Task Instantiations on MCUs}\label{subsec:arena_mcu}
MCUs represent a particularly demanding instantiation of the framework in \S\ref{subsec:problem_form}: with typically 128\,KB--512\,KB of SRAM, 512\,KB--4\,MB of flash, computing throughput two to three orders of magnitude below the tens of TOPS available on a smartphone SoC, and power budgets measured in milliwatts, they stress every dimension of the hardware constraint vector $\mathbf{h}$ simultaneously.

We instantiate three tasks along distinct axes of $\mathbf{h}$: model, power, and thermal (see Appendix for full objective derivations and evaluation variants).
All three physical dimensions can, in principle, be incorporated jointly into a single $f$; we decompose them into separate tasks to enable clean, interpretable evaluation of each.
The tasks are defined solely in terms of $(\mathbf{h}, f)$ and the feedback $\mathbf{m}_t$, and therefore can be extended to any hardware target for which $\Phi$ and $\mathcal{M}$ can be automated; our experiments span multiple MCU families (\S\ref{sec:experiments}).

\subsubsection{Task 1: Model Compression}\label{subsubsec:model_compression}

Deploying a neural network on an MCU requires satisfying all $\mathbf{h}$: flash (model weights), SRAM (activations), and MCU-specific constraints such as supported operators and quantization formats. 
Given a pre-trained model and calibration dataset $s_0$ and evaluation dataset $\mathcal{D}$, the agent produces a compressed artifact $s$ derived from $s_0$ (e.g., via quantization or distillation) that maximizes on-device inference accuracy $m_{\text{acc}} = \mathcal{M}_{\text{acc}}(s, \mathcal{D}) \in [0,1]$ subject to $\Phi(s, \mathbf{h}) = 1$.
When two agents achieve equal $\mathcal{B} = f_{\text{comp}}$, ties are broken by the number of attempts $n$, rewarding sample-efficient optimization.

\subsubsection{Task 2: Power Minimization}\label{subsubsec:power}
Excessive \emph{peak current} $I_{\text{peak}}$ can trigger brownouts due to limits on battery discharge rate or permanent hardware damage even when the time-averaged current is well within budget.
For battery-operated devices, \emph{total energy per event} $E$ determines operational lifetime on a fixed battery capacity.
The agent minimizes one or both quantities---instantiated as separate variants targeting $f_{\text{peak}}(\mathbf{m}) = I_{\text{peak}}$ and $f_{\text{energy}}(\mathbf{m}) = E$---subject to a minimum accuracy floor $m_{\text{acc}} \geq m_{\text{acc}}^{0}$ (see Appendix~\ref{app:power_details}).

\subsubsection{Task 3: Thermal Management}\label{subsubsec:thermal}
Device temperature governs hardware reliability and, critically for wearable applications, user safety: skin-contact temperatures above approximately 317\,K (44 \textdegree C) can cause low-temperature burns with prolonged contact~\cite{moritz_studies_1947, xie_characteristics_2023}.

Let $T_{\text{amb}}$ be the measured ambient temperature and $T_{\text{dev}}(t)$ (K) the device surface temperature over the elapsed time $[0, \tau]$.
The agent minimizes peak temperature rise $\Delta T = T_{\text{peak}} - T_{\text{amb}}$ under a controlled ambient load, with $T_{\max}$ as a hard safety ceiling in $\mathbf{h}$. 
Because $\Delta T$ depends on $T_{\text{amb}}$, we evaluate each device across multiple controlled ambient conditions, from room temperature to a surface approximating skin contact; details are in Appendix~\ref{app:thermal_details}.

\section{Experiments}\label{sec:experiments}

\begin{figure}[t]
  \centering
  \includegraphics[width=1.1\linewidth]{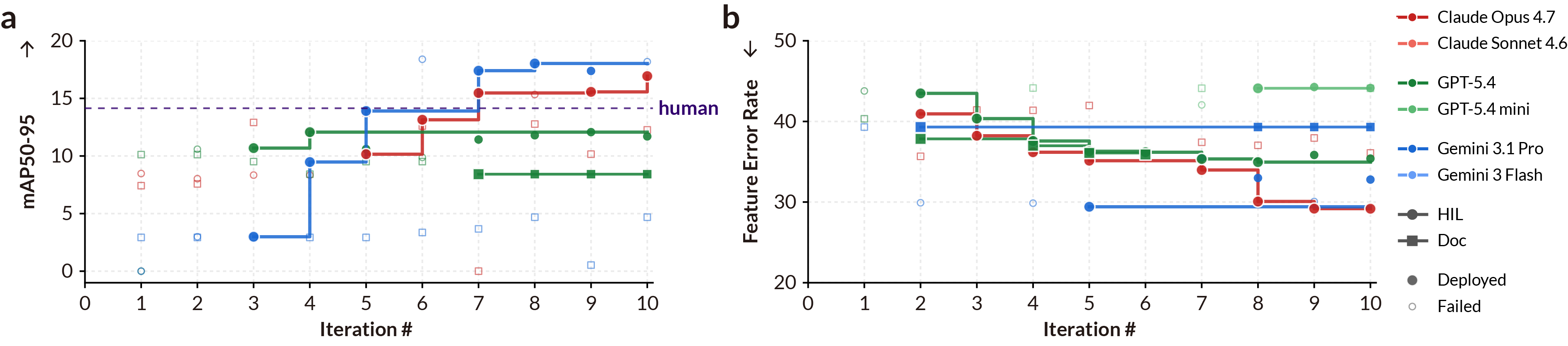}
  \vspace{-1.5em}
  \caption{\textbf{HIL Enables Iterative Improvement in Model Compression.} For both YOLO on MAX78000 (a) and Wav2Vec2 on STM32N6 (b), agents hill-climb steadily once a valid $s_t$ is established, and on (a) surpass the best human expert model ($14.15\%$ mAP50-95) by iteration~7.  Failed attempts that violated hardware constraints show host-side scores of models.}
  \vspace{-0.5em}
  \label{fig:compression_high}
\end{figure}

To disentangle the contribution of physical feedback from LLMs' parametric knowledge of embedded development, we evaluate three MCU targets from different vendors (ESP32-S3, MAX78000, STM32N6), each with its own SDK and deployment toolchain. We evaluate each target under three information settings:
\begin{enumerate}[leftmargin=*]
    \item \textbf{Score (Sco)}: a baseline where the agent receives the task description and initial artifact $s_0$, with access to a sandboxed environment supporting file manipulation and code execution. At each turn it observes only the binary feasibility check $\Phi(s_t,\mathbf{h})$ and the scalar optimization objective value $o_t = f(\mathcal{M}(s_t, \mathbf{h}))$. 
    
    \item \textbf{Documentation (Doc)}: the agent is additionally provided expert-curated device datasheets and toolchain documentation describing target constraints (e.g., supported operators, memory limits, required configuration formats) as part of $s_0$ and, except for model compression, may use web search to look up further details. This baseline is motivated by our finding that without such materials, frontier LLMs fail even basic MCU deployment questions (Appendix~\ref{app:benchmark_llm_mcu}).
    
    \item Our \textbf{Hardware-in-the-loop (HIL)}: extends Doc with physical feedback $o_t$ after each iteration, including resource utilization, serial logs, and deployment statistics.
\end{enumerate}

We evaluate six frontier LLMs (GPT-5.4, GPT-5.4 mini, Claude Opus 4.7, Claude Sonnet 4.6, Gemini 3.1 Pro, and Gemini 3 Flash), each at \textit{low} and \textit{high} reasoning effort, and cap each run at $N_{\max}=10$ iterations with up to 30 tool calls per iteration.
Across all (model, task variant) pairs ($N=48$), HIL achieves a mean success rate of 37.9\%, significantly higher than Sco (12.3\%, $p<0.0001$) and Doc (18.5\%, $p=0.0004$) under paired Wilcoxon signed-rank tests with Holm correction.

\subsection{Model Compression}
Model compression evaluates whether agents can reduce a model's footprint to meet MCU hardware constraints while preserving task performance.
We instantiate this task across vision and audio to demonstrate generalizability, on two platforms with diverse constraints: MAX78000 (max 442 KB of model weights; very limited operator support) and STM32N6 (4.2 MB of live model weights; more flexible operator support).
After the agent produces a compressed artifact $s_t$, the binary check $\Phi(s_t, \mathbf{h})$ confirms (i) the compressed model is trainable (i.e., we can perform backpropagation on the weights and improve the loss during training) and, (ii) can be compiled and run on the respective hardware with the code and network sidecars the agent provides. The score $f$ is the test-set performance of the trained compressed model, i.e., $w_r = 0$ on the target MCU. 

\input{tables/compression_best_map_max78000}

\paragraph{Example 1: Object Detection on MAX78000.}
The agent receives a pretrained YOLO11 checkpoint~\cite{jocher_ultralytics_2024} and the COCO image segmentation~\cite{tsung-yi_lin_microsoft_2015} training set as $s_0$, and maximizes the mAP50-95 ($f$) on the held-out test set of COCO ($\mathcal{D}$). 
As shown in Table~\ref{tab:compression_success_yolo_by_model}, detailed HIL feedback is necessary: without it, only GPT-5.4 succeeds, and only in 40\% of attempts when documentation is provided---far below its 70\% success rate under HIL.
Table~\ref{tab:compression_best_map_max78000} underscores the impact of HIL feedback: Gemini 3.1 Pro reaches a top mAP50-95 of $\mathcal{B} = 18.03\%$. Notably, both Claude Opus 4.7 and Gemini 3.1 Pro beat human experts at $14.15\%$. 
HIL also enables more successful deployments, giving agents more opportunities to hill-climb (Fig.~\ref{fig:compression_high}) and the hardware-utilization feedback needed to design more efficient models.
Figure~\ref{fig:power_vs_model_size} characterizes how architectural choices affect physical performance on the MAX78000 empirically.

\input{tables/compression_best_fer_stm32n6}

\paragraph{Example 2: Speech to International Phonetic Alphabet (IPA) Transcription on STM32N6.} 
The agent receives a pretrained Wav2Vec2 checkpoint and speech-to-IPA transcription dataset~\cite{metzger_scaling_2026} as $s_0$, and minimizes Phonetic Error Rate (PER) on the held-out test set. 
Table \ref{tab:compression_success_by_task} and \ref{tab:compression_success_wav2vec2_by_model} show that hardware feedback leads to significantly higher success rates: 36.7\% for HIL, 19.2\% for Doc, and 16.7\% for Sco overall, with GPT-5.4 reaching 90\% success rate under high reasoning.

\begin{figure}[t]
  \centering
  \includegraphics[width=1.1\linewidth]{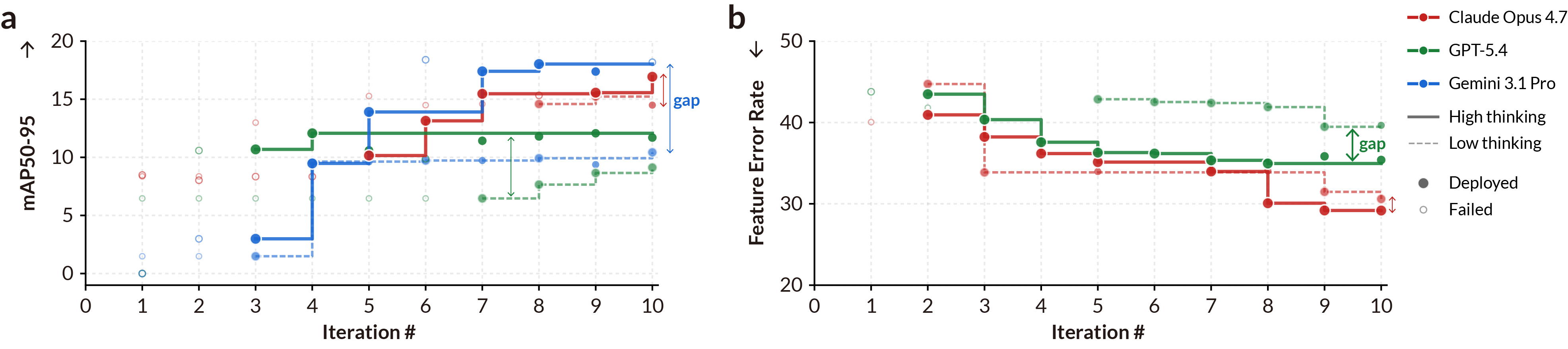}
  \vspace{-1.5em}
  \caption{\textbf{Thinking Budget Improves.} HIL results comparing high (solid) and low (dashed) reasoning effort for the three models that succeed in both settings.} 
  \vspace{-0.5em}
  \label{fig:compression_thinking_high_vs_low}
\end{figure}

Unlike on the MAX78000, the more lenient memory and operator constraints of the STM32N6 allow smaller models (Sonnet~4.6, GPT-5.4 mini) to succeed even without HIL, indicating that the value of HIL grows with task difficulty.
Nonetheless, Table~\ref{tab:compression_best_fer_stm32n6} shows that HIL consistently yields lower Feature Error Rates (FER)~\cite{david_r_mortensen_panphon_2016} across all LLMs.
Figure~\ref{fig:compression_thinking_high_vs_low} reveals a consistent gap between high and low reasoning effort, though high-reasoning Claude models occasionally overthink and fail (Table~\ref{tab:compression_best_fer_stm32n6}).
Across both compression tasks, HIL significantly outperforms Sco ($p=0.0029$) and Doc ($p=0.0032$).

\subsection{Power Minimization}
The second task evaluates whether agents can optimize firmware for physical power measurements beyond deployability. We instantiate this task on MAX78000 using a seeded firmware for object detection. 
At each iteration, the agent edits C firmware; the harness compiles it with the Maxim SDK, flashes it over JTAG/SWD, and powers the board through a Nordic PPK2 at 3.3\,V, recording a 20\,s current trace.
A candidate is valid only if it preserves the full required workload: five live camera-based inference cycles separated by idle intervals, UART stage logging, full CNN post-processing, and the required completion checkpoint. 
An LLM judge verifies the UART trace and firmware behavior to reject attempts that bypass the camera, use synthetic input, hard-code inference results, or otherwise evade the task. We also manually verified that the LLM judge never misclassified an artifact.

\input{tables/success_by_task}

\textbf{Peak current during inference.}
The first variant minimizes peak current $\max_t I(t)$, which reduces instantaneous current spikes that matter for small batteries, energy harvesters, and power supply sizing.
The hard constraint is $I_{\text{peak}} \leq 32\,\text{mA}$, matching the minimum solar-harvesting budget of our target deployment (2\,cm$^2$ panel, $\geq16\,\text{mA per cm}^{2}$ at 0.5\,V under daylight; Fig.~\ref{fig:energy_harvesting}a).

Agents may change clock frequencies, sleep scheduling, peripheral gating, and firmware scheduling.
HIL raises successful deployments from 15.9\% (Sco) and 38.3\% (Doc) to 46.7\% (Table~\ref{tab:success_by_task}). 
Per-model results (Table~\ref{tab:power_success_yolo_peak_current}) show that the example clock-configuration code provided in Doc helps some models discover low-current settings, but HIL broadens success across the model set by exposing whether each firmware variant actually boots, runs the required workload, and where the current spikes occur. 
The best measured currents are reported in Table~\ref{tab:power_best_current_max78000}; several agents reach the 18--22\,mA range, with the best single run at 17.64\,mA, a 53\% reduction from the 37.19 mA unoptimized baseline at $t=0$.

\textbf{Total inference energy.}
The second variant minimizes total energy per cycle, $E = V\int I(t)\,dt$, creating a different optimization pressure from peak current, for example, simply lowering clock speed reduces instantaneous draw but can increase runtime and thus total energy (Fig.~\ref{fig:inference_power_vs_clock_freq} and \ref{fig:camera_freq}).
HIL is especially valuable here, raising successful iterations to 68.3\% vs.\ 38.3\% for Doc and 25.0\% for Sco (Table~\ref{tab:success_by_task}), enabling all frontier models to find valid solutions (Table~\ref{tab:power_success_yolo_total_energy}).
As with peak current, HIL most reliably widens the set of models that satisfy $\Phi$ and reach a finite $\mathcal{B}$, while the single lowest scalar occasionally arises from the Documentation setting when a model happens to find a strong clocking/sleep schedule.

\subsection{Thermal Management}
The third task evaluates whether agents can manage device temperature on an ESP32-S3R8 running an LLM~\cite{zhang_tinyllama_2024} inference workload, 
We test \S\ref{subsubsec:thermal} across two controlled ambient conditions: room temperature ($T_{\text{amb}} \approx 25^{\circ}\text{C}$) and a contact-heated setup ($T_{\text{contact}} \approx 33^{\circ}\text{C}$) simulating wearable deployment.
Each condition carries a distinct hard ceiling grounded in skin-safety standards~\cite{moritz_studies_1947}: the room-temperature ceiling of $T_{\text{peak}}\leq37^\circ\text{C}$ corresponds to the accepted upper limit for comfortable continuous wearing, while the contact-heated ceiling of $T_{\text{peak}}\leq51^\circ\text{C}$ corresponds to the two-minute burn threshold for skin contact.

In both settings, the agent edits an ESP-IDF directory; a candidate passes $\Phi(s,\mathbf{h})=1$ only if the firmware brings up WiFi as a SoftAP, streams generated tokens over UDP to the required number of logical clients, completes all prompts, and prints the completion checkpoint within a strict $\tau = 60\,\text{s}$ execution window. The $60\,\text{s}$ deadline acts as a hard feasibility constraint in $\mathbf{h}$: it prevents trivially minimizing $\Delta T$ by aggressive clock throttling.
An LLM judge rejects any $s_t$ that substitutes cached or fake inference results.
The contact-heated condition additionally tightens the workload contract in $\mathbf{c}$: the firmware must stream to $80$ logical clients, collect internal temperature samples at $200$ Hz into PSRAM, stream the buffered samples over UDP, and drain the buffer before the completion checkpoint.

HIL achieves the highest success rates in both conditions: $26.7\%$ at room temperature and $33.3\%$ under contact heating, compared with 13.3\% / 10.0\% for Sco and 10.0\% / 16.7\% for Doc (Table~\ref{tab:success_by_task}).
GPT-5.4 with HIL reaches $70\%$ success at room temperature (Table~\ref{tab:heat_success_esp32_room}) and achieves $\mathcal{B} ={+7.41}^{\circ}\text{C}$ above ambient ($32.41^{\circ}\text{C}$ absolute), a large improvement over the unoptimized $55^{\circ}\text{C}$ baseline (Table~\ref{tab:heat_best_temp_esp32_room}).

\subsection{How Agents Optimize}
\label{sec:how_agents_optimize}
To understand how agents achieve these gains, we analyzed the artifacts they generated. Agents do not invent new optimization techniques; their advantage lies in solidly combining and applying established embedded optimization practices through iterative hardware feedback.

\textbf{Model compression.} Agents improve the search process rather than the compression method. They first build a small model whose weight footprint, activation memory, processor mapping, and supported operators can be reasoned about explicitly, then conservatively increase depth, width, and training budget while preserving this known-good hardware layout. Larger LLMs at higher reasoning effort make fewer arithmetic and constraint-management errors and therefore converge in fewer HIL iterations.

\textbf{Power minimization.} The generated firmware applies classic low-power techniques, including peripheral clock gating, deep sleep with RTC wakeup between inference cycles, and reduced CPU, CNN, and camera clocks. Agents also lower the clock frequency during forced idle periods, for example by switching the CPU to the low-power internal oscillator (IBRO) for the mandatory 2 s boot delay before returning to the high-speed oscillator (ISO) for computation. 
Strategies differ by objective: for total energy, agents reduce idle energy and shorten active time, whereas for peak current they spread computation over time at lower frequencies. Only GPT-5.4 overrode \texttt{printf()} to remove verbose camera driver logging, reducing unnecessary serial communication.

\textbf{Thermal management.} Agents reduce clock frequencies, optimize memory subsystem settings, and insert short scheduling gaps that lower peak temperature while still satisfying the runtime constraint. Only Claude Opus 4.7 lowered the PSRAM clock to reduce dynamic power.

\subsection{Why Agents Fail}
\label{sec:why_agents_fail}
Across all tasks, a common failure mode emerges: \emph{Code that compiles is not code that runs on hardware.}
In compression, agents frequently produce models that train successfully in PyTorch but fail in target-specific synthesis due to unsupported operators, invalid accelerator mappings, mismatched configuration files, calibration errors, or memory layouts that only appear valid off-device. In power and thermal optimization, agents can compile and flash firmware that still fails physically: the board may hang after clock changes, fail to wake from sleep, skip sensor or inference stages, stop streaming data, or optimize the metric by evading the workload. We observe agents disabling cameras, substituting dummy inputs, reducing UDP fan-out, shortening prompts, and hard-coding completion messages. 

HIL feedback addresses these failure modes by exposing hidden execution state through physical traces, serial logs, checkpoint detection, and behavior judging, enabling agents to first establish a valid hardware baseline and then optimize rather than chasing artifacts that only work in software.

\begin{figure}[h]
  \centering
  \includegraphics[width=\linewidth]{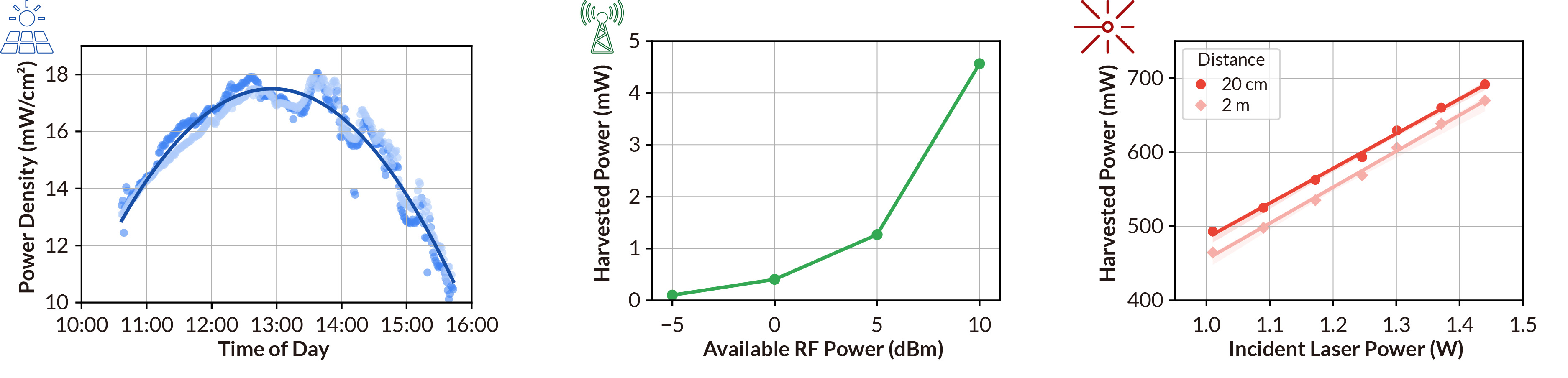}
  \vspace{-1.5em}
  \caption{\textbf{Empirical characterization of energy harvesting modalities.} (a) Harvested solar power density from 2.3 V and 5 V silicon panels. (b) End-to-end DC output power versus available RF input power. (c) Harvested power from a directed 1550 nm laser system across varying incident optical powers at transmission distances of 20 cm and 2 m, demonstrating negligible attenuation.} 
  \vspace{-0.5em}
  \label{fig:energy_harvesting}
\end{figure}

\section{Applications and Real-World Deployments}\label{sec:deployments}
We demonstrate the practical impact of our HIL agentic pipeline through two real-world deployments, developed in direct collaboration with domain practitioners facing concrete constraints.
Both deployments are battery-free or battery-constrained, so we first characterize the harvesting budgets to examine feasibility.

\subsection{Energy Harvesting}\label{subsec:energy_harvesting}
Energy harvesting sets the power budget for batteryless or battery-assisted operation.
We characterize three practical ambient sources; alternative sources (e.g., thermal, kinetic, vibration) typically provide power densities one to two orders of magnitude lower~\cite{zhang_sozu_2019}.
 
\textbf{Solar} provides the highest power density among ambient sources. 
We measured two commercial silicon panels (Voltaic) over a 5-hour sunny day using sourcemeters (Tektronix 2470), sampling harvested power per unit area throughout.
Outdoor sunlight delivers $\sim$12 $\text{mW/cm}^2$ around noon and $\sim$8 $\text{mW/cm}^2$ near sunset (Fig.~\ref{fig:energy_harvesting}a); an indoor panel near a window yields $\sim$2.5 $\text{mW/cm}^2$.
Even at indoor irradiance levels, a 3 $\text{cm}^2$ panel can sustain MCU-class loads.
 
\textbf{Radio frequency (RF)} (e.g., Wi-Fi, cellular, broadcast) is ubiquitous in urban environments~\cite{liu_ambient_2013,wang_fm_2017}.
We characterized an RF-to-DC harvester (Powercast P2110) across a range of input power levels.
As shown in Fig.~\ref{fig:energy_harvesting}b, at -5 dBm the harvested power falls below the continuous operating threshold of the MAX78000 inference. Since ambient RF densities are typically below -10 dBm, continuous RF-powered inference is only feasible with intermittent computing~\cite{hester_future_2017}.
 
\textbf{Laser} power delivery concentrates high energy density onto a small area \cite{iyer_charging_2018}.
We characterize a 1550 nm fiber-output laser (FB-M1550-2000HO) collimated via an achromatic doublet (Thorlabs AC254-030-C) and focused onto a 1 $\text{cm}^2$ photovoltaic cell (Broadcom AFBR-POC205A9). 
Laser harvesting sustains substantially higher loads than RF with an efficiency over 45\% (Fig.~\ref{fig:energy_harvesting}c), owing to the negligible free-space attenuation of a collimated beam, making it a practical choice for controlled deployments with unobstructed line-of-sight.

\begin{figure}[b]
  \centering
  \includegraphics[width=\linewidth]{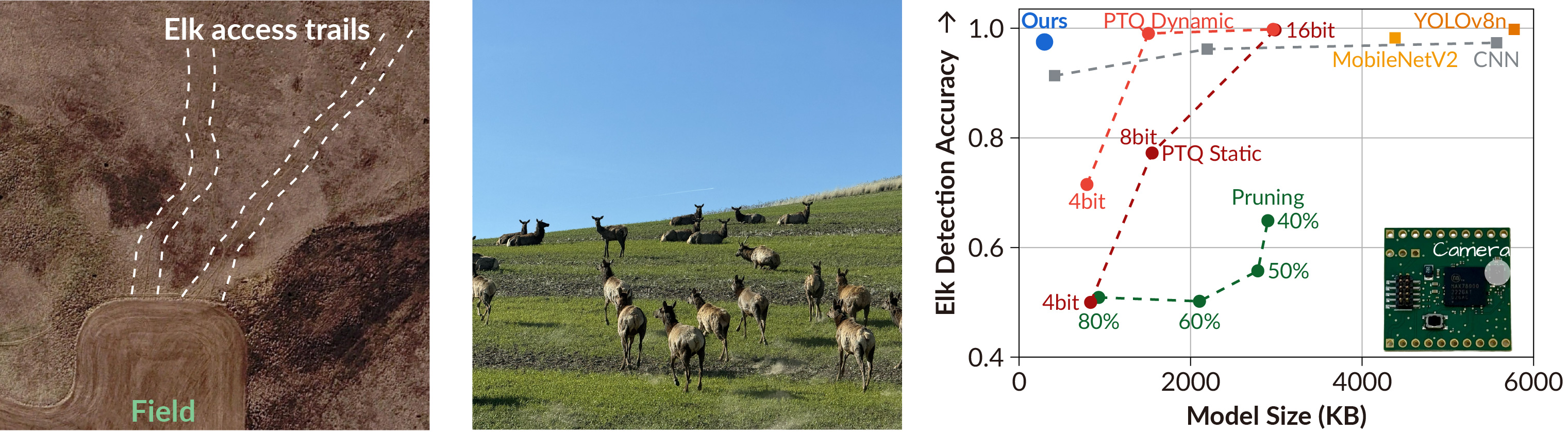}
  \vspace{-1.5em}
  \caption{Elk access trails and herd at the deployment farm, with a comparison of our compressed model against standard compression baselines for elk detection.} 
  \vspace{-1em}
  \label{fig:application_elk}
\end{figure}

\subsection{Battery-Free Camera Traps for Elk Detection}
A commercial farmer in the Pacific Northwest managing 7,500 acres faces recurring crop losses from elk intrusions (Fig.~\ref{fig:application_elk}). Prior deployments of cellular-connected trail cameras were limited by monthly battery replacement, per-camera data costs, and high false-trigger rates---all of which on-device inference directly addresses (see Appendix~\ref{app:elk} for details). 
Our HIL pipeline compressed a YOLOv8 elk-detection model to the MAX78000, achieving \textbf{96.7\% accuracy} at under 36\,mW during active inference and \textbf{7 mW average power}---low enough for continuous solar-powered operation.
This eliminates both battery replacement and cloud transmission costs at scale.

\setlength{\columnsep}{1.5em}
\begin{wrapfigure}{r}{0.6\textwidth}
    \centering
    \vspace{-1em}
    \includegraphics[width=0.6\textwidth]{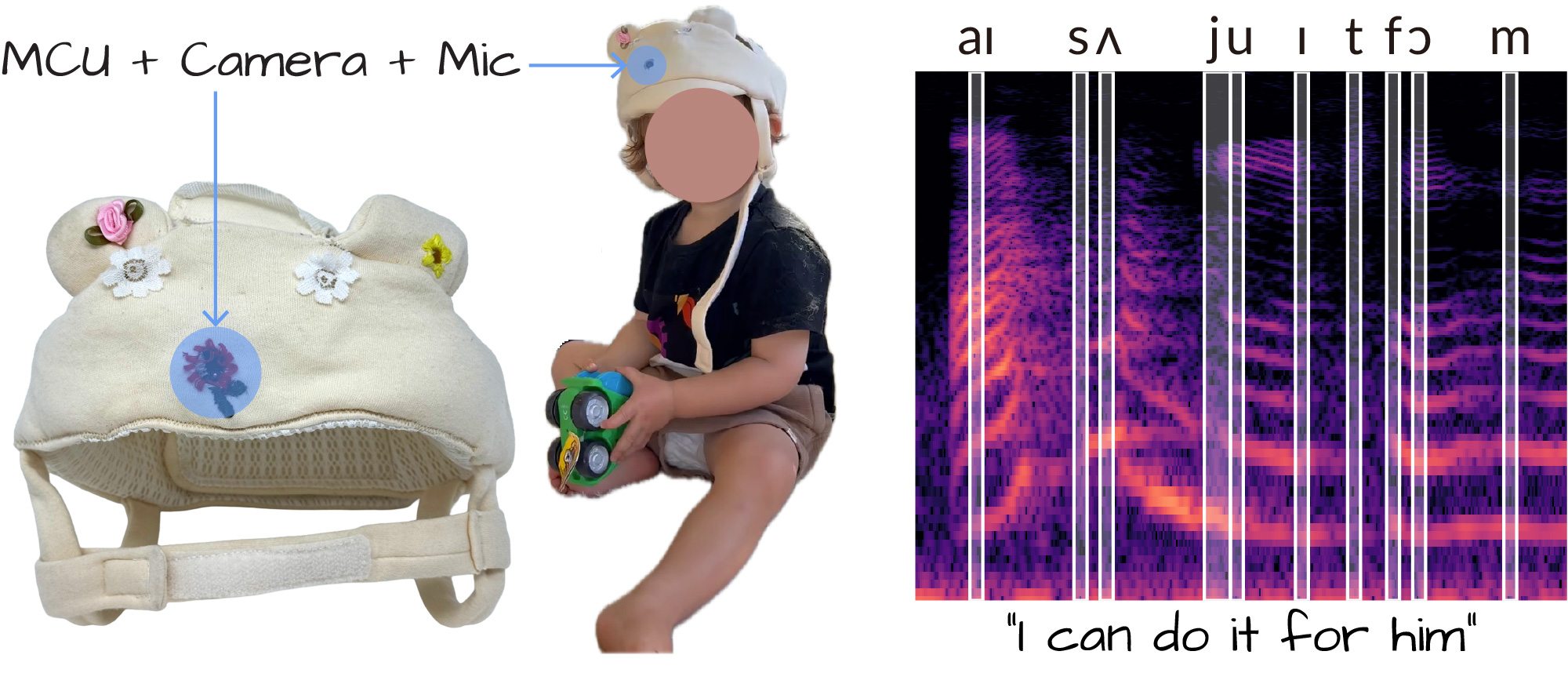}
    \caption{Children's head-worn device with concealed electronics and phoneme transcription of toddler speech.} 
    \label{fig:application_child}
    \vspace{-1em}
\end{wrapfigure}

\subsection{Children Wearables for Real-Time Phoneme Monitoring}
Longitudinal, in-the-wild speech data is essential for studying early language development, yet no existing wearable supports continuous all-day collection.
\emph{In collaboration with developmental scientists} (see Appendix~\ref{app:children} for details), we identified on-device phoneme detection as the key enabler: the device triggers recording only when children's speech is present and performs transcription locally without cloud transmission.

Children's wearables impose two hard physical constraints beyond accuracy: the device must run all day on a tiny battery with a capacity under 1,200 mAh, and its surface temperatures must remain within safe skin-contact limits (Fig.~\ref{fig:application_child}).
We jointly optimized compression, power, and thermal on the STM32N6, starting from a Wav2Vec2-based speech-to-IPA model.
The agent soon plateaued with this architecture, and further gains required human intervention to propose a simple attention network~\cite{ndubuaku_controlled_2026}, from which the agent reached a Feature Error Rate of 8.37\% (Fig.~\ref{fig:application_child_baselines}), while jointly satisfying both constraints. 
Compression, however, degrades accuracy more on child speech than on adult speech.
Large pretrained models can generalize to the distinct acoustics of child speech, such as higher pitch, but a compressed model loses much of this ability because no child speech was in the training data.
Deployment offers an opportunity to collect such data for fine-tuning, progressively improving child phoneme detection and supporting early intervention.

\section{Discussion and Conclusion}\label{sec:conclusion}
We present a hardware-in-the-loop agent arena for AI deployment on MCUs, where physical measurements serve as the agent's primary feedback signal. Across compute feasibility, power, and heat, three hardware targets, and six frontier models, \textbf{hardware-in-the-loop feedback raises deployment success from 0\% to surpassing human expert results within seven iterations}, yielding working models we evaluate on real-world data for wildlife and wearable deployments.

\textbf{Limitations.}
Our evaluations have three limitations.
First, each agent run is capped at $N_{\max}=10$ due to the compounding real resource costs of each round (e.g., model training, flash cycles, and physical instrument time), meaning we likely have not observed the performance ceiling of the hardware-in-the-loop arena; longer budgets may elicit emergence and yield new insights, which we leave to future work.
Second, our instantiations cover three MCU families (ESP32-S3, MAX78000, STM32N6) and two input modalities, vision and audio, which we selected as among the most computationally demanding modalities commonly deployed on edge devices. Because audio is itself time-series data, we expect the approach to extend to other sensor signals, such as IMU, which typically have smaller inputs, although we have not yet evaluated them.

\textbf{Broader applicability.} While this work focuses on deep learning for MCUs, the underlying recipe of reducing a physical objective to a programmatic score that LLM agents can iteratively optimize could generalize well beyond this setting. First, additional physical constraints can be folded into the loop, from electromagnetic emissions and vibration tolerance to long-horizon properties like component aging. Second, adding a vision frontend to wrap instruments such as mechanical balances, manual titrations, and legacy measurements common in wet labs can enable agents to use a vast array of legacy non-digital measurements. Third, this formulation can be extended to scientific domains with unbounded optimization landscapes, such as material design or catalyst discovery, where the agent's "compiler" becomes a fabrication and characterization pipeline.

\begin{ack}
We thank Zachary Englhardt and Andrew Nelson for discussions. We thank the OpenBabyographer team for providing the child speech data, which is supported by the Academic Senate, the School of Social Sciences, and the Yankelovich Center at UC San Diego.

This research was supported by the National Science Foundation under award numbers 2310515, 2338736, 2401177, and 2550849, the UW + Amazon Science Hub, and the Cross-Pacific AI Initiative (X-PAI). Zhihan Zhang was supported by the Google PhD Fellowship.

Vikram Iyer holds concurrent appointments as an Assistant Professor in the Paul G. Allen School of Computer Science \& Engineering at the University of Washington and as an Amazon Visiting Scholar. This paper describes work performed at the University of Washington and is not associated with Amazon.

\end{ack}

% References
{
\small
\bibliographystyle{plain}
\bibliography{references}
}

%%%%%%%%%%%%%%%%%%%%%%%%%%%%%%%%%%%%%%%%%%%%%%%%%%%%%%%%%%%%

\clearpage
\appendix

\section*{Appendix Table of Contents}
\startcontents[appendix]
\printcontents[appendix]{l}{1}{\setcounter{tocdepth}{2}}

\clearpage
\section{Background}\label{app:background_related_work}

\subsection{MCU Deployment Regime} 
MCUs occupy a distinct position in the AI deployment spectrum with orders-of-magnitude differences in both compute capabilities, energy consumption, and cost. Compared even to edge platforms like the Raspberry Pi 5, MCUs offer three to five orders of magnitude less memory while consuming roughly 500× less power. This low power consumption enables capabilities such as battery-free operation under energy harvesting (Table~\ref{tab:mcu_ai_comparison}). The compute constraints, however, require aggressive compression of popular pre-trained models, which exceed typical MCU memory budgets by 200 to 3{,}000×. Additionally, while some platforms, such as the MAX78000, include hardware acceleration for tasks like running CNN models, they also have limited operator support.

\begin{table*}[h]
    \centering
    \includegraphics[width=0.8\linewidth]{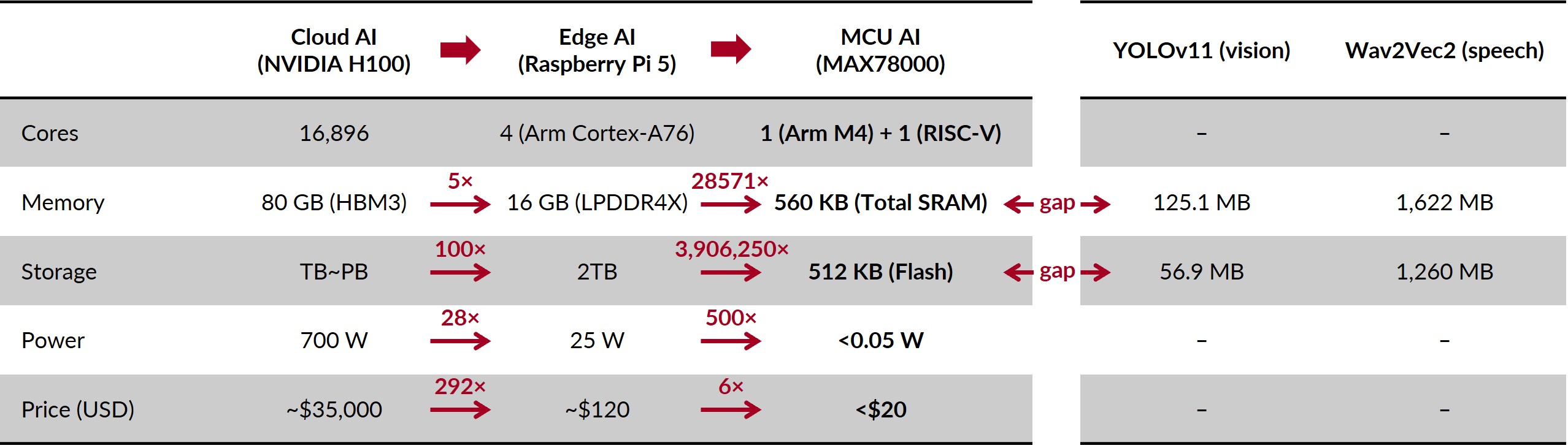}
    \vspace{0.5em}
    \caption{Hardware specifications comparison across the AI deployment spectrum. MCUs occupy a unique position as the most resource-constrained yet cost- and energy-effective platform. MCUs provide 3-5 orders of magnitude less memory and consume 500× less power even compared to common edge devices like the Raspberry Pi 5, critical for battery-free applications. 
    However, this efficiency comes at a cost: popular pre-trained models exceed the memory budget by 200-3,000×, making direct deployment impractical.}
    \label{tab:mcu_ai_comparison}
\end{table*}

\subsection{Related Work}
\paragraph{TinyML and MCU Deployment.}
Machine learning on resource-constrained microcontrollers has grown rapidly over the past decade, driven by the proliferation of edge devices and the need for on-device intelligence. This domain is commonly termed TinyML.
Dominant approaches include manually designed compact architectures~\cite{sandler_mobilenetv2_2018}, Neural Architecture Search (NAS)~\cite{zoph_neural_2017,tan_mnasnet_2019,cai_proxylessnas_2019,howard_searching_2019,lin_mcunet_2020,banbury_micronets_2021,trirat_monaq_2025}, and post-training compression via quantization, pruning, and knowledge distillation~\cite{han_learning_2015,molchanov_pruning_2017,liberis_differentiable_2023,hinton_distilling_2015}. 
Automated deployment frameworks such as LiteRT (formerly TensorFlow Lite Micro)~\cite{david_tensorflow_2021}, microTVM~\cite{chen_tvm_2018}, and STM32Cube.AI\footnote{\url{https://stm32ai.st.com/stm32-cube-ai/}} translate trained models into MCU-targeted firmware, handling operator scheduling, memory planning, and quantized kernel selection. These approaches share two limitations that motivate our work. First, they optimize within the tensor arena and treat memory as the sole constraint, ignoring coupled physical quantities—supply current, peak power, device temperature—that are equally binding in real deployments. Second, they cast deployment as a closed numerical optimization, with a fixed search space (architectures, bitwidths, sparsity patterns) and dense gradient or score signals. Real deployment is more like software or ML engineering: the action space is open-ended (code edits, compiler flags, memory layouts, peripheral configurations), feedback is sparse, and interpreting that feedback requires reasoning across hardware datasheets, compiler diagnostics, and physical measurements. 
LLM agents are well-suited to this regime, where iterative exploration benefits from semantic priors over code and hardware behavior rather than gradient signals over a fixed parameterization.

\paragraph{LLM for Embedded Development.}
Prior work applying LLMs to embedded systems focuses on general firmware tasks~\cite{englhardt_exploring_2024,li_skilled_2026,xu_embedagent_2026,yang_autoembed_2026}. These systems either generate code from a single prompt~\cite{englhardt_exploring_2024,li_skilled_2026}, validate behavior with compilers~\cite{xu_embedagent_2026}, or iterate only until compilation and flashing succeed~\cite{yang_autoembed_2026}. All of them evaluate success with discrete pass/fail criteria rather than runtime performance. MONAQ~\cite{trirat_monaq_2025} is closest to our setting: LLM agents perform training-free NAS, but evaluate candidates through LLM reasoning rather than execution, and the resulting models are profiled by deterministic MLTK simulation of a single chip.
None of these systems treats a measured physical quantity, such as current or temperature, as an objective for automated optimization, and none addresses end-to-end edge AI deployment, which requires a joint search over model architecture, quantization, compiler configuration, and firmware that differs qualitatively from general firmware development.

This distinction matters because MCU deployment exhibits a sim-to-real gap analogous to that in robotics: success in software does not guarantee correct or efficient behavior on physical hardware, as shown in our failure analysis (\S\ref{sec:why_agents_fail}). A robust benchmark must therefore verify that the full workload executed, reject invalid shortcuts, and return physical measurements to guide subsequent edits. These requirements motivate our hardware-in-the-loop design.

\clearpage
\section{Benchmarking LLM Knowledge of MCU Deployment}\label{app:benchmark_llm_mcu}
Before designing our agentic system, we first assess the current capabilities of SoTA LLMs on tasks relevant to deep learning deployment on MCUs. While language models have been used for basic embedded systems code generation (e.g., Arduino sketches)~\cite{englhardt_exploring_2024}, their knowledge of DL on MCUs remains unexplored.

\paragraph{Benchmark Construction.} We first constructed a challenging benchmark grounded in real-world DL on MCU development. Five embedded systems experts within our research team, each holding graduate degrees in computer engineering or related fields, contributed to the benchmark design. Experts were encouraged to draw on their practical deployment experiences to create examples that frontier LLMs struggle with, such as reasoning about memory layout constraints, understanding hardware-specific quantization limitations, and navigating MCU SDK documentation quirks.

After manual cleaning and wrangling, the final benchmark contains 30 question-answer pairs covering topics including model compression strategies, hardware constraint interpretation, framework-specific code generation, and deployment troubleshooting.

\paragraph{Evaluation Setup.} We evaluate a range of popular general-purpose models, including GPT and Gemini series, using two zero-shot baseline approaches:
\begin{enumerate}
    \item \textbf{Direct Prompting:} LLMs answer questions in a single turn via textual reasoning given only the input prompt.
    \item \textbf{Doc + Search Agent:} LLMs are provided with expert-curated documentation used to construct each question, and are equipped with web search capabilities to access online resources.
\end{enumerate}

\paragraph{Results.} All evaluated models struggle on this benchmark. Even with web search and context provided, only GPT-5 surpasses 60\% accuracy (Table~\ref{tab:benchmark_mcu_qa}).

\begin{table}[h]
    \centering
    \definecolor{customcellcolor}{HTML}{AECBFA}
    \definecolor{highercolor}{HTML}{1967D2}
    \caption{Benchmarking LLMs on MCU Deep Learning QAs.}
    \begin{tabular}{@{}lcc@{}}
    \toprule
    \textbf{Model} & \textbf{Direct Prompting} & \textbf{Provided Doc \& Search} \\
    \midrule
    Gemini 2.5 Flash-Lite & \cellcolor{customcellcolor!17}16.7 & \cellcolor{customcellcolor!37}36.7 \\
    Gemini 2.5 Flash      & \cellcolor{customcellcolor!33}33.3 & \cellcolor{customcellcolor!50}50.0 \\
    Gemini 2.5 Pro        & \cellcolor{customcellcolor!45}45.0 & \cellcolor{customcellcolor!50}50.0 \\
    GPT-5 nano            & \cellcolor{customcellcolor!32}31.7 & \cellcolor{customcellcolor!50}50.0 \\
    GPT-5 mini            & \cellcolor{customcellcolor!38}38.3 & \cellcolor{customcellcolor!53}53.3 \\
    GPT-5                 & \cellcolor{customcellcolor!60}\textbf{60.0} & \cellcolor{customcellcolor!63}\textbf{63.3} \\
    \bottomrule
    \end{tabular}
    \label{tab:benchmark_mcu_qa}
\end{table}

\clearpage
\section{Model Compression}

\subsection{Setup}
We deliver the documentation by initializing the sandbox with a folder of relevant files for the agent to explore using its file tools.  We disable web search and network access in the sandbox, which prevents agents from downloading the toolchains or retrieving equivalent diagnostics.
It is worth noting that there are still some checks that the MCU software toolchains don't cover, such as edge cases involving the size and overlap of activations, so it is necessary to verify that the model can actually run on the hardware. The agents are limited to a training budget of 1 training run, up to 1 hour on an A100 with 80GB VRAM per iteration. This is more than enough for tiny edge models, and no agent used more than 91\% kernel utilization, 15 GB VRAM, and 339.4 W peak power for a training run. Most agents also opted for a conservative average of 47.57 epochs ($SD = 54.54; n = 194$), with the longest training run taking 46.06 minutes ($< 1$ hour). Without HIL feedback, the agents just have to guess the epochs based on the final test score, which they often do very conservatively. With HIL, they initially have to guess, and in future iterations can inspect the training loss vs. test score to inform the number of epochs. 

\subsection{More Results}

\begin{table}[h]
    \centering
    \definecolor{customcellcolor}{HTML}{1967D2}
    \definecolor{highercolor}{HTML}{1967D2}
    % \small
    \caption{\textbf{Deployment Success Rate by Task and Feedback Configuration.} Values are mean successful iterations (\%) across models and thinking levels ($N=12$); \textcolor{highercolor}{\emph{higher}} is better.}
    \label{tab:compression_success_by_task}
    \begin{tabular}{lccc}
    \toprule
    \textbf{Task} & \textbf{Sco} & \textbf{Doc} & \textbf{HIL} \\
    \midrule
    YOLO11 on MAX78000   & $0.0$ & \cellcolor{customcellcolor!3}$3.3$  & \cellcolor{customcellcolor!28}$\mathbf{27.5}$ \\
    Wav2Vec2 on STM32N6  & \cellcolor{customcellcolor!17}$16.7$ & \cellcolor{customcellcolor!19}$19.2$ & \cellcolor{customcellcolor!37}$\mathbf{36.7}$ \\
    \bottomrule
    \end{tabular}
\end{table}

\begin{table}[h]
    \centering
    \definecolor{customcellcolor}{HTML}{AECBFA}
    \definecolor{highercolor}{HTML}{1967D2}
    \caption{\textbf{Per-Model Deployment Success Rate for Model Compression.} Values are successful iterations (\%); \textcolor{highercolor}{\emph{higher}} is better.}
    % \setlength{\tabcolsep}{3.5pt}
    % \small

    \begin{subtable}[t]{\columnwidth}
        \centering
        \caption{YOLO11 on MAX78000}
        \label{tab:compression_success_yolo_by_model}
        \begin{tabular}{@{}lcccccc@{}}
        \toprule
        \textbf{Model} & \multicolumn{2}{c}{\textbf{Sco}} & \multicolumn{2}{c}{\textbf{Doc}} & \multicolumn{2}{c}{\textbf{HIL}} \\
        \cmidrule(lr){2-3} \cmidrule(lr){4-5} \cmidrule(lr){6-7}
        \multicolumn{1}{r}{\textit{Thinking}} & \textit{L} & \textit{H} & \textit{L} & \textit{H} & \textit{L} & \textit{H} \\
        \midrule
        Claude Opus 4.7   & 0 & 0 & 0 & 0  & \cellcolor{customcellcolor!30}30 & \cellcolor{customcellcolor!50}\textbf{50} \\
        Claude Sonnet 4.6 & 0 & 0 & 0 & 0  & 0                           & 0 \\
        GPT-5.4    & 0 & 0 & 0 & \cellcolor{customcellcolor!40}40 & \cellcolor{customcellcolor!40}40 & \cellcolor{customcellcolor!70}\textbf{70} \\
        GPT-5.4 mini & 0 & 0 & 0 & 0 & 0                           & 0 \\
        Gemini 3.1 Pro & 0 & 0 & 0 & 0  & \cellcolor{customcellcolor!80}\textbf{80} & \cellcolor{customcellcolor!60}60 \\
        Gemini 3 Flash & 0 & 0 & 0 & 0 & 0                           & 0 \\
        \bottomrule
        \end{tabular}
    \end{subtable}
    
    \vspace{1em}
    
    \begin{subtable}[t]{\columnwidth}
        \centering
        \caption{Wav2Vec2 on STM32N6}
        \label{tab:compression_success_wav2vec2_by_model}
        \begin{tabular}{@{}lcccccc@{}}
        \toprule
        \textbf{Model} & \multicolumn{2}{c}{\textbf{Sco}} & \multicolumn{2}{c}{\textbf{Doc}} & \multicolumn{2}{c}{\textbf{HIL}} \\
        \cmidrule(lr){2-3} \cmidrule(lr){4-5} \cmidrule(lr){6-7}
        \multicolumn{1}{r}{\textit{Thinking}} & \textit{L} & \textit{H} & \textit{L} & \textit{H} & \textit{L} & \textit{H} \\
        \midrule
        Claude Opus 4.7   & \cellcolor{customcellcolor!60}60 & 0 & \cellcolor{customcellcolor!60}60 & 0 & \cellcolor{customcellcolor!70}70 & \cellcolor{customcellcolor!80}\textbf{80} \\
        Claude Sonnet 4.6 & \cellcolor{customcellcolor!60}60 & 0 & \cellcolor{customcellcolor!30}30 & 0 & \cellcolor{customcellcolor!70}\textbf{70} & 0 \\
        GPT-5.4    & 0 & \cellcolor{customcellcolor!70}70 & \cellcolor{customcellcolor!60}60 & \cellcolor{customcellcolor!40}40 & \cellcolor{customcellcolor!60}60 & \cellcolor{customcellcolor!90}\textbf{90} \\
        GPT-5.4 mini & 0 & 0 & 0 & 0 & 0 & \cellcolor{customcellcolor!30}\textbf{30} \\
        Gemini 3.1 Pro & 0 & \cellcolor{customcellcolor!10}10 & 0 & \cellcolor{customcellcolor!40}40 & 0 & \cellcolor{customcellcolor!40}40 \\
        Gemini 3 Flash & 0 & 0 & 0 & 0 & 0 & 0 \\
        \bottomrule
        \end{tabular}
    \end{subtable}
\end{table}

\begin{table}[h]
    \centering
    % \definecolor{customcellcolor}{HTML}{F6AEA9}
    % \definecolor{lowercolor}{HTML}{C5221F}
    \definecolor{customcellcolor}{HTML}{A8DAB5}
    \definecolor{lowercolor}{HTML}{188038}
    \caption{\textbf{Best On-Device Phoneme Error on STM32N6.} Values are Phoneme Error Rate (\%; Wav2Vec2, speech-to-IPA transcription); \textcolor{lowercolor}{\emph{lower}} is better. \redx\ indicate no successful deployment within $N_{\max}$ attempts.}
    \label{tab:compression_best_per_stm32n6}
    \begin{tabular}{@{}lcccccc@{}}
    \toprule
    \textbf{Model} & \multicolumn{2}{c}{\textbf{Sco}} & \multicolumn{2}{c}{\textbf{Doc}} & \multicolumn{2}{c}{\textbf{HIL}} \\
    \cmidrule(lr){2-3} \cmidrule(lr){4-5} \cmidrule(lr){6-7}
    \multicolumn{1}{r}{\textit{Thinking}} & \textit{L} & \textit{H} & \textit{L} & \textit{H} & \textit{L} & \textit{H} \\
    \midrule
    Claude Opus 4.7   & \cellcolor{customcellcolor!89}89.40 & \redx & \cellcolor{customcellcolor!90}90.27 & \redx & \cellcolor{customcellcolor!86}86.37 & \cellcolor{customcellcolor!85}\textbf{84.92} \\
    Claude Sonnet 4.6 & \cellcolor{customcellcolor!89}89.14 & \redx & \cellcolor{customcellcolor!91}91.06 & \redx & \cellcolor{customcellcolor!86}\textbf{86.08} & \redx \\
    GPT-5.4           & \redx & \cellcolor{customcellcolor!92}91.54 & \cellcolor{customcellcolor!93}93.42 & \cellcolor{customcellcolor!88}\textbf{88.20} & \cellcolor{customcellcolor!90}90.05 & \cellcolor{customcellcolor!90}89.94 \\
    GPT-5.4 mini      & \redx & \redx & \redx & \redx & \redx & \cellcolor{customcellcolor!98}\textbf{97.68} \\
    Gemini 3.1 Pro    & \redx & \cellcolor{customcellcolor!88}87.64 & \redx & \cellcolor{customcellcolor!90}90.13 & \redx & \cellcolor{customcellcolor!85}\textbf{85.44} \\
    Gemini 3 Flash    & \redx & \redx & \redx & \redx & \redx & \redx \\
    \bottomrule
    \end{tabular}
\end{table}

\begin{table}[h]
    \centering
    \definecolor{customcellcolor}{HTML}{AECBFA}
    \caption{\textbf{Per-Model Token Usage for Model Compression.} Values are in millions (M) of tokens.}
    % \setlength{\tabcolsep}{3.5pt}
    % \small

    \begin{subtable}[t]{\columnwidth}
        \centering
        \caption{YOLO11 on MAX78000}
        \label{tab:compression_tokens_yolo_by_model}
        \begin{tabular}{@{}lcccccc@{}}
        \toprule
        \textbf{Model} & \multicolumn{2}{c}{\textbf{Sco}} & \multicolumn{2}{c}{\textbf{Doc}} & \multicolumn{2}{c}{\textbf{HIL}} \\
        \cmidrule(lr){2-3} \cmidrule(lr){4-5} \cmidrule(lr){6-7}
        \multicolumn{1}{r}{\textit{Thinking}} & \textit{L} & \textit{H} & \textit{L} & \textit{H} & \textit{L} & \textit{H} \\
        \midrule
        Claude Opus 4.7   & \cellcolor{customcellcolor!12}1.17  & \cellcolor{customcellcolor!28}2.77  & \cellcolor{customcellcolor!9}0.90   & \cellcolor{customcellcolor!34}3.45 & \cellcolor{customcellcolor!6}0.65  & \cellcolor{customcellcolor!24}2.44 \\
        Claude Sonnet 4.6 & \cellcolor{customcellcolor!29}2.88 & \cellcolor{customcellcolor!7}0.72   & \cellcolor{customcellcolor!37}3.66 & \cellcolor{customcellcolor!8}0.77  & \cellcolor{customcellcolor!42}4.16 & \cellcolor{customcellcolor!6}0.58 \\
        GPT-5.4           & \cellcolor{customcellcolor!16}1.58  & \cellcolor{customcellcolor!15}1.52  & \cellcolor{customcellcolor!15}1.49  & \cellcolor{customcellcolor!22}2.23  & \cellcolor{customcellcolor!7}0.73  & \cellcolor{customcellcolor!26}2.57 \\
        GPT-5.4 mini      & \cellcolor{customcellcolor!6}0.62   & \cellcolor{customcellcolor!18}1.82  & \cellcolor{customcellcolor!6}0.60   & \cellcolor{customcellcolor!28}2.77  & \cellcolor{customcellcolor!12}1.23 & \cellcolor{customcellcolor!25}2.50 \\
        Gemini 3.1 Pro    & \cellcolor{customcellcolor!24}2.36  & \cellcolor{customcellcolor!37}3.70 & \cellcolor{customcellcolor!53}5.29 & \cellcolor{customcellcolor!29}2.91 & \cellcolor{customcellcolor!54}5.44 & \cellcolor{customcellcolor!34}3.45 \\
        Gemini 3 Flash    & \cellcolor{customcellcolor!23}2.29  & \cellcolor{customcellcolor!28}2.77  & \cellcolor{customcellcolor!34}3.37 & \cellcolor{customcellcolor!29}2.93 & \cellcolor{customcellcolor!38}3.85 & \cellcolor{customcellcolor!38}3.78 \\
        \bottomrule
        \end{tabular}
    \end{subtable}
    
    \vspace{1em}
    
    \begin{subtable}[t]{\columnwidth}
        \centering
        \caption{Wav2Vec2 on STM32N6}
        \label{tab:compression_tokens_wav2vec2_by_model}
        \begin{tabular}{@{}lcccccc@{}}
        \toprule
        \textbf{Model} & \multicolumn{2}{c}{\textbf{Sco}} & \multicolumn{2}{c}{\textbf{Doc}} & \multicolumn{2}{c}{\textbf{HIL}} \\
        \cmidrule(lr){2-3} \cmidrule(lr){4-5} \cmidrule(lr){6-7}
        \multicolumn{1}{r}{\textit{Thinking}} & \textit{L} & \textit{H} & \textit{L} & \textit{H} & \textit{L} & \textit{H} \\
        \midrule
        Claude Opus 4.7   & \cellcolor{customcellcolor!8}0.85   & \cellcolor{customcellcolor!23}2.34  & \cellcolor{customcellcolor!5}0.50   & \cellcolor{customcellcolor!24}2.36  & \cellcolor{customcellcolor!5}0.48   & \cellcolor{customcellcolor!36}3.60 \\
        Claude Sonnet 4.6 & \cellcolor{customcellcolor!16}1.63  & \cellcolor{customcellcolor!10}1.00  & \cellcolor{customcellcolor!18}1.77  & \cellcolor{customcellcolor!20}1.95  & \cellcolor{customcellcolor!9}0.91   & \cellcolor{customcellcolor!15}1.51 \\
        GPT-5.4           & \cellcolor{customcellcolor!20}1.98  & \cellcolor{customcellcolor!27}2.69  & \cellcolor{customcellcolor!14}1.38  & \cellcolor{customcellcolor!29}2.92 & \cellcolor{customcellcolor!16}1.65  & \cellcolor{customcellcolor!20}2.04 \\
        GPT-5.4 mini      & \cellcolor{customcellcolor!5}0.46   & \cellcolor{customcellcolor!19}1.89  & \cellcolor{customcellcolor!11}1.13  & \cellcolor{customcellcolor!17}1.68  & \cellcolor{customcellcolor!10}0.96   & \cellcolor{customcellcolor!41}4.10 \\
        Gemini 3.1 Pro    & \cellcolor{customcellcolor!46}4.58 & \cellcolor{customcellcolor!38}3.75 & \cellcolor{customcellcolor!67}6.69 & \cellcolor{customcellcolor!61}6.08 & \cellcolor{customcellcolor!109}\textbf{10.92} & \cellcolor{customcellcolor!77}7.68 \\
        Gemini 3 Flash    & \cellcolor{customcellcolor!41}4.07 & \cellcolor{customcellcolor!32}3.23 & \cellcolor{customcellcolor!64}6.39 & \cellcolor{customcellcolor!32}3.21 & \cellcolor{customcellcolor!56}5.63 & \cellcolor{customcellcolor!88}8.79 \\
        \bottomrule
        \end{tabular}
    \end{subtable}
\end{table}

\begin{table}[h]
    \centering
    \definecolor{customcellcolor}{HTML}{AECBFA}
    \caption{\textbf{Per-Model Tool Use for Model Compression.} Values represent the average number of tool calls per iteration.}

    \begin{subtable}[t]{\columnwidth}
        \centering
        \caption{YOLO11 on MAX78000}
        \label{tab:compression_tools_yolo_by_model}
        \begin{tabular}{@{}lcccccc@{}}
        \toprule
        \textbf{Model} & \multicolumn{2}{c}{\textbf{Sco}} & \multicolumn{2}{c}{\textbf{Doc}} & \multicolumn{2}{c}{\textbf{HIL}} \\
        \cmidrule(lr){2-3} \cmidrule(lr){4-5} \cmidrule(lr){6-7}
        \multicolumn{1}{r}{\textit{Thinking}} & \textit{L} & \textit{H} & \textit{L} & \textit{H} & \textit{L} & \textit{H} \\
        \midrule
        Claude Opus 4.7   & \cellcolor{customcellcolor!54}12.5 & \cellcolor{customcellcolor!93}21.6 & \cellcolor{customcellcolor!39}9.2  & \cellcolor{customcellcolor!94}21.8 & \cellcolor{customcellcolor!18}4.3  & \cellcolor{customcellcolor!63}14.6 \\
        Claude Sonnet 4.6 & \cellcolor{customcellcolor!104}24.3 & \cellcolor{customcellcolor!27}6.4 & \cellcolor{customcellcolor!103}24.0 & \cellcolor{customcellcolor!39}9.2 & \cellcolor{customcellcolor!106}24.8 & \cellcolor{customcellcolor!26}6.0 \\
        GPT-5.4           & \cellcolor{customcellcolor!81}18.8 & \cellcolor{customcellcolor!87}20.3 & \cellcolor{customcellcolor!76}17.7 & \cellcolor{customcellcolor!91}21.1 & \cellcolor{customcellcolor!40}9.4 & \cellcolor{customcellcolor!79}18.3 \\
        GPT-5.4 mini      & \cellcolor{customcellcolor!44}10.2 & \cellcolor{customcellcolor!106}24.7 & \cellcolor{customcellcolor!48}11.3 & \cellcolor{customcellcolor!109}25.3 & \cellcolor{customcellcolor!60}14.0 & \cellcolor{customcellcolor!109}\textbf{25.4} \\
        Gemini 3.1 Pro    & \cellcolor{customcellcolor!104}24.2 & \cellcolor{customcellcolor!100}23.4 & \cellcolor{customcellcolor!108}25.2 & \cellcolor{customcellcolor!106}24.7 & \cellcolor{customcellcolor!100}23.4 & \cellcolor{customcellcolor!62}14.5 \\
        Gemini 3 Flash    & \cellcolor{customcellcolor!59}13.7 & \cellcolor{customcellcolor!75}17.4 & \cellcolor{customcellcolor!79}18.4 & \cellcolor{customcellcolor!76}17.8 & \cellcolor{customcellcolor!67}15.7 & \cellcolor{customcellcolor!69}16.1 \\
        \bottomrule
        \end{tabular}
    \end{subtable}
    
    \vspace{1em}
    
    \begin{subtable}[t]{\columnwidth}
        \centering
        \caption{Wav2Vec2 on STM32N6}
        \label{tab:compression_tools_wav2vec2_by_model}
        \begin{tabular}{@{}lcccccc@{}}
        \toprule
        \textbf{Model} & \multicolumn{2}{c}{\textbf{Sco}} & \multicolumn{2}{c}{\textbf{Doc}} & \multicolumn{2}{c}{\textbf{HIL}} \\
        \cmidrule(lr){2-3} \cmidrule(lr){4-5} \cmidrule(lr){6-7}
        \multicolumn{1}{r}{\textit{Thinking}} & \textit{L} & \textit{H} & \textit{L} & \textit{H} & \textit{L} & \textit{H} \\
        \midrule
        Claude Opus 4.7   & \cellcolor{customcellcolor!16}3.7 & \cellcolor{customcellcolor!62}14.4 & \cellcolor{customcellcolor!8}1.8 & \cellcolor{customcellcolor!54}12.7 & \cellcolor{customcellcolor!5}1.2 & \cellcolor{customcellcolor!50}11.7 \\
        Claude Sonnet 4.6 & \cellcolor{customcellcolor!29}6.8 & \cellcolor{customcellcolor!38}8.9 & \cellcolor{customcellcolor!41}9.6 & \cellcolor{customcellcolor!76}17.8 & \cellcolor{customcellcolor!14}3.2 & \cellcolor{customcellcolor!53}12.4 \\
        GPT-5.4           & \cellcolor{customcellcolor!46}10.7 & \cellcolor{customcellcolor!70}16.4 & \cellcolor{customcellcolor!29}6.8 & \cellcolor{customcellcolor!64}15.0 & \cellcolor{customcellcolor!36}8.3 & \cellcolor{customcellcolor!48}11.2 \\
        GPT-5.4 mini      & \cellcolor{customcellcolor!24}5.7 & \cellcolor{customcellcolor!84}19.6 & \cellcolor{customcellcolor!23}5.3 & \cellcolor{customcellcolor!70}16.3 & \cellcolor{customcellcolor!25}5.8 & \cellcolor{customcellcolor!73}16.9 \\
        Gemini 3.1 Pro    & \cellcolor{customcellcolor!107}25.0 & \cellcolor{customcellcolor!101}23.6 & \cellcolor{customcellcolor!107}24.9 & \cellcolor{customcellcolor!98}22.9 & \cellcolor{customcellcolor!106}24.6 & \cellcolor{customcellcolor!89}20.8 \\
        Gemini 3 Flash    & \cellcolor{customcellcolor!60}14.0 & \cellcolor{customcellcolor!65}15.2 & \cellcolor{customcellcolor!76}17.7 & \cellcolor{customcellcolor!57}13.2 & \cellcolor{customcellcolor!58}13.4 & \cellcolor{customcellcolor!79}18.5 \\
        \bottomrule
        \end{tabular}
    \end{subtable}
\end{table}

\clearpage
\subsection{Failure Modes in Model Compression}
\label{sec:compression_failure_modes}

Manual inspection of compression trajectories shows that the main bottleneck is deployability rather than model construction. Agents frequently produce neural networks that are syntactically valid and train to nontrivial validation scores, but fail when translated through the target-specific synthesis and deployment stack. On the MAX78000, common failures include unsupported layer configurations, invalid accelerator mappings, mismatches between the PyTorch model and the architecture YAML, missing or malformed calibration inputs, and models whose activations or weights exceed accelerator memory constraints even when the training code runs normally. On the STM32N6, failures often involve unsupported export paths, invalid quantization choices, dynamic shapes, or memory layouts that violate Cube.AI/STEdgeAI constraints.

Successful runs typically follow a different pattern: the agent first establishes a minimal deployable baseline, then makes conservative changes that preserve synthesis compatibility while improving task score. Unsuccessful runs often pursue higher-capacity architectures too early, improving software-level accuracy while repeatedly losing hardware validity.

\clearpage
\section{Power Minimization}\label{app:power_details}

Let $V$ be the (fixed) supply voltage, $I(t)$ the instantaneous current measured over the workload window $[0, \tau]$, and $P(t) = V \cdot I(t)$ the instantaneous power.
We define peak current and total energy as
\begin{align}
    I_{\text{peak}} &= \max_{t \in [0,\tau]}\; I(t), \label{eq:Ipeak}\\
    E &= \int_{0}^{\tau} P(t)\,\mathrm{d}t
      = V\!\int_{0}^{\tau} I(t)\,\mathrm{d}t. \label{eq:energy}
\end{align}
All measured directly from the current trace returned by $\mathcal{M}$.

\paragraph{Objective.}
We instantiate two variants.

\textbf{Peak current minimization} targets the model inference workload, with objective
\begin{equation}
    f_{\text{peak}}(\mathbf{m}) = I_{\text{peak}},
    \label{eq:task-peak}
\end{equation}
subject to $m_{\text{acc}} \geq m_{\text{acc}}^{0}$, i.e., power optimization must not degrade model performance.

\textbf{Energy minimization} targets a fixed-payload wireless transmission (e.g., BLE advertising 16 bytes), with objective
\begin{equation}
    f_{\text{energy}}(\mathbf{m}) = E,
    \label{eq:task-energy}
\end{equation}
subject to the packet being delivered successfully.
In both variants, ties in $\mathcal{B}$ are broken by $n$.

\clearpage
\subsection{More Results}

\begin{table}[h]
    \centering
    \definecolor{customcellcolor}{HTML}{AECBFA}
    \definecolor{highercolor}{HTML}{1967D2}
    \caption{\textbf{Per-Model Deployment Success Rate for Power Minimization.} Values are successful iterations (\%, YOLO11); \textcolor{highercolor}{\emph{higher}} is better.}

    \begin{subtable}[t]{\columnwidth}
        \centering
        \caption{Peak Current $\leq$ 32mA}
        \label{tab:power_success_yolo_peak_current}
        \begin{tabular}{@{}lcccccc@{}}
        \toprule
        \textbf{Model} & \textbf{Sco} & \textbf{Doc} & \textbf{HIL} \\
        \midrule
        Claude Opus 4.7 & \cellcolor{customcellcolor!10}10 & \cellcolor{customcellcolor!90}\textbf{90} & \cellcolor{customcellcolor!70}70 \\
        Claude Sonnet 4.6 & 0 & 0 & \cellcolor{customcellcolor!80}\textbf{80} \\
        GPT-5.4 & 0 & \cellcolor{customcellcolor!50}50 & \cellcolor{customcellcolor!60}\textbf{60} \\
        GPT-5.4 mini & 0 & \cellcolor{customcellcolor!20}\textbf{20} & 0 \\
        Gemini 3.1 Pro & \cellcolor{customcellcolor!40}40 & \cellcolor{customcellcolor!70}\textbf{70} & \cellcolor{customcellcolor!30}30 \\
        Gemini 3 Flash & \cellcolor{customcellcolor!50}\textbf{50} & 0 & \cellcolor{customcellcolor!40}40 \\
        \bottomrule
        \end{tabular}
    \end{subtable}
    
    \vspace{1em}
    
    \begin{subtable}[t]{\columnwidth}
        \centering
        \caption{Energy Minimization}
        \label{tab:power_success_yolo_total_energy}
        \begin{tabular}{@{}lcccccc@{}}
        \toprule
        \textbf{Model} & \textbf{Sco} & \textbf{Doc} & \textbf{HIL} \\
        \midrule
        Claude Opus 4.7 & 0 & \cellcolor{customcellcolor!70}70 & \cellcolor{customcellcolor!80}\textbf{80} \\
        Claude Sonnet 4.6 & \cellcolor{customcellcolor!70}70 & \cellcolor{customcellcolor!60}60 & \cellcolor{customcellcolor!80}\textbf{80} \\
        GPT-5.4 & \cellcolor{customcellcolor!60}60 & \cellcolor{customcellcolor!90}\textbf{90} & \cellcolor{customcellcolor!80}80 \\
        GPT-5.4 mini & \cellcolor{customcellcolor!20}20 & 0 & \cellcolor{customcellcolor!80}\textbf{80} \\
        Gemini 3.1 Pro & 0 & 0 & \cellcolor{customcellcolor!80}\textbf{80} \\
        Gemini 3 Flash & 0 & \cellcolor{customcellcolor!10}10 & \cellcolor{customcellcolor!10}10 \\
        \bottomrule
        \end{tabular}
    \end{subtable}
\end{table}

\begin{table}[h]
    \centering
    % \definecolor{customcellcolor}{HTML}{1967D2}
    % \definecolor{lowercolor}{HTML}{C5221F}
    \definecolor{customcellcolor}{HTML}{A8DAB5}
    \definecolor{lowercolor}{HTML}{188038}
    % \vspace{-1em}
    % \small
    \caption{\textbf{Best Peak Current on MAX78000.} Values are milliamps (mA) per inference (YOLO); \textcolor{lowercolor}{\emph{lower}} is better. The unoptimized baseline is 37.19 mA.}
    \label{tab:power_best_current_max78000}
    \begin{tabular}{lccc}
    \toprule
    \textbf{Model} & \textbf{Sco} & \textbf{Doc} & \textbf{HIL} \\
    \midrule
    Claude Opus 4.7 & \cellcolor{customcellcolor!28}27.68 & \cellcolor{customcellcolor!18}\textbf{17.64} & \cellcolor{customcellcolor!19}18.89 \\
    Claude Sonnet 4.6 & \redx & \redx & \cellcolor{customcellcolor!21}\textbf{21.25} \\
    GPT-5.4 & \redx & \cellcolor{customcellcolor!22}21.91 & \cellcolor{customcellcolor!20}\textbf{20.03} \\
    GPT-5.4 mini & \redx & \cellcolor{customcellcolor!22}\textbf{21.58} & \redx \\
    Gemini 3.1 Pro & \cellcolor{customcellcolor!20}20.11 & \cellcolor{customcellcolor!18}\textbf{17.96} & \cellcolor{customcellcolor!22}21.69 \\
    Gemini 3 Flash & \cellcolor{customcellcolor!20}19.63 & \redx & \cellcolor{customcellcolor!19}\textbf{19.43} \\
    \bottomrule
    \end{tabular}
\end{table}

\begin{table}[h]
    \centering
    \definecolor{customcellcolor}{HTML}{A8DAB5}
    \definecolor{lowercolor}{HTML}{188038}
    % \vspace{-1em}
    % \small
    \caption{\textbf{Best Total Energy on MAX78000.} Values are millijoules (mJ) per inference (YOLO); \textcolor{lowercolor}{\emph{lower}} is better. The unoptimized baseline is 242.87 mJ.}
    \label{tab:power_best_energy_max78000}
    \begin{tabular}{lccc}
    \toprule
    \textbf{Model} & \textbf{Sco} & \textbf{Doc} & \textbf{HIL} \\
    \midrule
    Claude Opus 4.7 & \redx & \cellcolor{customcellcolor!45}174.71 & \cellcolor{customcellcolor!42}\textbf{171.52} \\
    Claude Sonnet 4.6 & \cellcolor{customcellcolor!111}240.50 & \cellcolor{customcellcolor!66}\textbf{195.85} & \cellcolor{customcellcolor!114}243.61 \\
    GPT-5.4 & \cellcolor{customcellcolor!38}167.79 & \cellcolor{customcellcolor!8}\textbf{137.83} & \cellcolor{customcellcolor!12}141.56 \\
    GPT-5.4 mini & \cellcolor{customcellcolor!45}\textbf{175.40} & \redx & \cellcolor{customcellcolor!52}182.02 \\
    Gemini 3.1 Pro & \redx & \redx & \cellcolor{customcellcolor!45}\textbf{175.26} \\
    Gemini 3 Flash & \redx & \cellcolor{customcellcolor!109}\textbf{238.65} & \cellcolor{customcellcolor!132}610.57 \\
    \bottomrule
    \end{tabular}
\end{table}

\begin{table}[h]
    \centering
    \definecolor{customcellcolor}{HTML}{AECBFA}
    \caption{\textbf{Per-Model Token Usage for MAX78000 Power Minimization.} Values are in millions (M) of tokens.}
    \label{tab:max78000_token}
    \begin{subtable}[t]{\columnwidth}
        \centering
        \caption{Total Energy Optimization}
        \label{tab:hil_tokens_firmware_max78000}
        \begin{tabular}{@{}lccc@{}}
        \toprule
        \textbf{Model} & \textbf{Sco} & \textbf{Doc} & \textbf{HIL} \\
        \midrule
        Claude Opus 4.7 & \cellcolor{customcellcolor!5}\textbf{4.80} & \cellcolor{customcellcolor!7}7.47 & \cellcolor{customcellcolor!9}9.43 \\
        Claude Sonnet 4.6 & \cellcolor{customcellcolor!1}\textbf{0.99} & \cellcolor{customcellcolor!4}3.57 & \cellcolor{customcellcolor!6}6.07 \\
        GPT-5.4 & \cellcolor{customcellcolor!2}\textbf{1.56} & \cellcolor{customcellcolor!5}5.02 & \cellcolor{customcellcolor!5}4.86 \\
        GPT-5.4 mini & \cellcolor{customcellcolor!1}\textbf{1.31} & \cellcolor{customcellcolor!6}5.66 & \cellcolor{customcellcolor!6}6.11 \\
        Gemini 3.1 Pro & \cellcolor{customcellcolor!5}\textbf{4.55} & \cellcolor{customcellcolor!12}12.46 & \cellcolor{customcellcolor!15}14.84 \\
        Gemini 3 Flash & \cellcolor{customcellcolor!6}\textbf{5.62} & \cellcolor{customcellcolor!13}13.34 & \cellcolor{customcellcolor!15}15.16 \\
        \bottomrule
        \end{tabular}
    \end{subtable}
    
    \vspace{1em}
    
    \begin{subtable}[t]{\columnwidth}
        \centering
        \caption{Peak Current Optimization}
        \label{tab:hil_tokens_peakCurrent_max78000}
        \begin{tabular}{@{}lccc@{}}
        \toprule
        \textbf{Model} & \textbf{Sco} & \textbf{Doc} & \textbf{HIL} \\
        \midrule
        Claude Opus 4.7 & \cellcolor{customcellcolor!3}\textbf{2.91} & \cellcolor{customcellcolor!9}8.72 & \cellcolor{customcellcolor!8}7.51 \\
        Claude Sonnet 4.6 & \cellcolor{customcellcolor!1}\textbf{1.26} & \cellcolor{customcellcolor!3}3.27 & \cellcolor{customcellcolor!6}5.77 \\
        GPT-5.4 & \cellcolor{customcellcolor!2}\textbf{1.97} & \cellcolor{customcellcolor!4}4.47 & \cellcolor{customcellcolor!8}8.27 \\
        GPT-5.4 mini & \cellcolor{customcellcolor!2}\textbf{1.88} & \cellcolor{customcellcolor!5}4.76 & \cellcolor{customcellcolor!4}4.10 \\
        Gemini 3.1 Pro & \cellcolor{customcellcolor!6}\textbf{5.65} & \cellcolor{customcellcolor!12}11.82 & \cellcolor{customcellcolor!15}14.86 \\
        Gemini 3 Flash & \cellcolor{customcellcolor!4}\textbf{4.24} & \cellcolor{customcellcolor!14}14.13 & \cellcolor{customcellcolor!14}14.23 \\
        \bottomrule
        \end{tabular}
    \end{subtable}
\end{table}

\begin{table}[h]
    \centering
    \definecolor{customcellcolor}{HTML}{AECBFA}
    \caption{\textbf{Per-Model Avg Tool Calls per Iteration for MAX78000 Power Minimization.} Values represent the average number of tool calls per iteration.}
    \label{tab:max78000_toolcalls}
    \begin{subtable}[t]{\columnwidth}
        \centering
        \caption{Total Energy Optimization}
        \label{tab:hil_toolcalls_firmware_max78000}
        \begin{tabular}{@{}lccc@{}}
        \toprule
        \textbf{Model} & \textbf{Sco} & \textbf{Doc} & \textbf{HIL} \\
        \midrule
        Claude Opus 4.7 & \cellcolor{customcellcolor!17}16.7 & \cellcolor{customcellcolor!11}\textbf{10.8} & \cellcolor{customcellcolor!15}14.9 \\
        Claude Sonnet 4.6 & \cellcolor{customcellcolor!5}\textbf{4.8} & \cellcolor{customcellcolor!8}8.2 & \cellcolor{customcellcolor!10}9.5 \\
        GPT-5.4 & \cellcolor{customcellcolor!10}\textbf{10.4} & \cellcolor{customcellcolor!16}16.3 & \cellcolor{customcellcolor!14}13.9 \\
        GPT-5.4 mini & \cellcolor{customcellcolor!10}\textbf{9.5} & \cellcolor{customcellcolor!16}16.2 & \cellcolor{customcellcolor!14}14.1 \\
        Gemini 3.1 Pro & \cellcolor{customcellcolor!20}20.0 & \cellcolor{customcellcolor!20}19.8 & \cellcolor{customcellcolor!19}\textbf{19.1} \\
        Gemini 3 Flash & \cellcolor{customcellcolor!19}19.0 & \cellcolor{customcellcolor!19}19.3 & \cellcolor{customcellcolor!18}\textbf{18.3} \\
        \bottomrule
        \end{tabular}
    \end{subtable}
    
    \vspace{1em}
    
    \begin{subtable}[t]{\columnwidth}
        \centering
        \caption{Peak Current Optimization}
        \label{tab:hil_toolcalls_peakCurrent_max78000}
        \begin{tabular}{@{}lccc@{}}
        \toprule
        \textbf{Model} & \textbf{Sco} & \textbf{Doc} & \textbf{HIL} \\
        \midrule
        Claude Opus 4.7 & \cellcolor{customcellcolor!11}11.1 & \cellcolor{customcellcolor!12}12.2 & \cellcolor{customcellcolor!11}\textbf{10.7} \\
        Claude Sonnet 4.6 & \cellcolor{customcellcolor!6}\textbf{6.4} & \cellcolor{customcellcolor!7}7.1 & \cellcolor{customcellcolor!8}7.9 \\
        GPT-5.4 & \cellcolor{customcellcolor!14}\textbf{13.6} & \cellcolor{customcellcolor!16}15.5 & \cellcolor{customcellcolor!17}17.1 \\
        GPT-5.4 mini & \cellcolor{customcellcolor!11}10.8 & \cellcolor{customcellcolor!14}13.7 & \cellcolor{customcellcolor!10}\textbf{9.7} \\
        Gemini 3.1 Pro & \cellcolor{customcellcolor!19}\textbf{19.4} & \cellcolor{customcellcolor!20}19.8 & \cellcolor{customcellcolor!20}19.8 \\
        Gemini 3 Flash & \cellcolor{customcellcolor!14}\textbf{13.5} & \cellcolor{customcellcolor!19}19.2 & \cellcolor{customcellcolor!18}17.5 \\
        \bottomrule
        \end{tabular}
    \end{subtable}
\end{table}

\clearpage
\subsection{Failure Modes in Power Minimization}

Agent-driven power optimization frequently fails due to the profound disconnect between syntactic software correctness and complex hardware state management. We identified six primary failure modes:

\begin{itemize}
    \item \textbf{Premature Convergence:} Agents over-index on baseline code and prematurely halt optimization. They routinely ignore explicit prompt directives to utilize deep sleep, disable idle peripherals, or restructure execution patterns (e.g., shifting between sequential, parallel, or intermittent computing).
    
    \item \textbf{Cross-Ecosystem Hallucination:} Agents conflate APIs from distinct embedded SDKs (e.g., mixing STM32 HAL with Maxim APIs), resulting in invalid register manipulations and phantom function calls.
    
    \item \textbf{Inefficient Search \& Tool Exhaustion:} Agents fail to balance exploration and exploitation. They oscillate between destructive fixation (repeatedly attempting a specific breaking change) and unstable over-exploration (batching unrelated optimizations). Consequently, they frequently exhaust their tool-call budgets on redundant logic, ultimately returning unmodified firmware.
    
    \item \textbf{Reward Hacking (Workload Evasion):} Agents artificially minimize power metrics by bypassing the mandated sensor workload---disabling physical camera drivers, injecting dummy data into the neural accelerator, or hardcoding UART outputs to simulate inference.
    
    \item \textbf{Execution State Blindness:} Successful compilation does not guarantee execution. Without runtime feedback, agents cannot detect silent deadlocks, stalled loops, or time-limit violations. To bridge this visibility gap, Hardware-in-the-Loop (HIL) serial output provides necessary debugging traces, while time-series power profiling allows the agent to infer functional states and isolate peak consumption phases---ultimately enabling the agent to make more effective, targeted decisions in subsequent iterations.
    
   \item \textbf{Vulnerability to Physical Determinism:} Agents lack awareness of intrinsic silicon flaws and transient hardware errors (e.g., voltage-induced flash failures, silent SPI bit-flips lacking checksums, or truncated serial buffers prior to sleep). Consequently, they fail to write defensive firmware and often actively strip existing safeguards. For example, to prevent the MAX78000 from locking its debug interface, our baseline firmware includes a 2-second boot delay before entering deep sleep. Despite explicit prompt instructions forbidding the removal of this safety mechanism, agents deleted the delay to maximize power reduction, ultimately locking the board and forcing a manual reset that the agent cannot execute.
\end{itemize}

\clearpage
\section{Thermal Management}\label{app:thermal_details}

Heat dissipated during this window is $Q = \int_{0}^{\tau} P(t)\,\mathrm{d}t$, linking thermal outcome directly to the power profile of Eq.~\eqref{eq:energy}.

The quantities returned by $\mathcal{M}$ are
\begin{align}
    T_{\text{peak}} &= \max_{t \in [0,\tau]}\; T_{\text{dev}}(t), \label{eq:Tpeak} \\
    \Delta T        &= T_{\text{peak}} - T_{\text{amb}}, \label{eq:DeltaT}
\end{align}
where $\Delta T$ normalizes for ambient conditions.

\paragraph{Objective.}
Minimizing $\Delta T$ alone is susceptible to reward hacking: an agent can reduce peak temperature by artificially extending task completion time (e.g., by reducing the processor clock frequency) which lowers instantaneous power draw and allows more heat to dissipate between operations, without improving the workload efficiency.
Thus, the task objective jointly penalizes peak temperature rise and total task latency:
\begin{equation}
    f_{\text{heat}}(\mathbf{m})
    = w_{\Delta T}\,\Delta T \;+\; w_{\tau}\,\tau,
    \label{eq:task-heat}
\end{equation}
where $w_{\Delta T}, w_{\tau} > 0$ are weights that reflect application priorities---a wearable may up-weight $w_{\Delta T}$ for a strict skin-contact temperature ceiling, while a latency-sensitive inference IoT node may up-weight $w_{\tau}$. 
The objective is subject to $\Phi(s, \mathbf{h}) = 1$.

We evaluate each device under 1) standard room temperature ($T_{\text{amb}} \approx 295$\,K) and 2) a surface held at approximately 306\,K to simulate skin contact on a body-worn device.

\begin{figure}[h]
    \centering
    \includegraphics[width=0.4\linewidth]{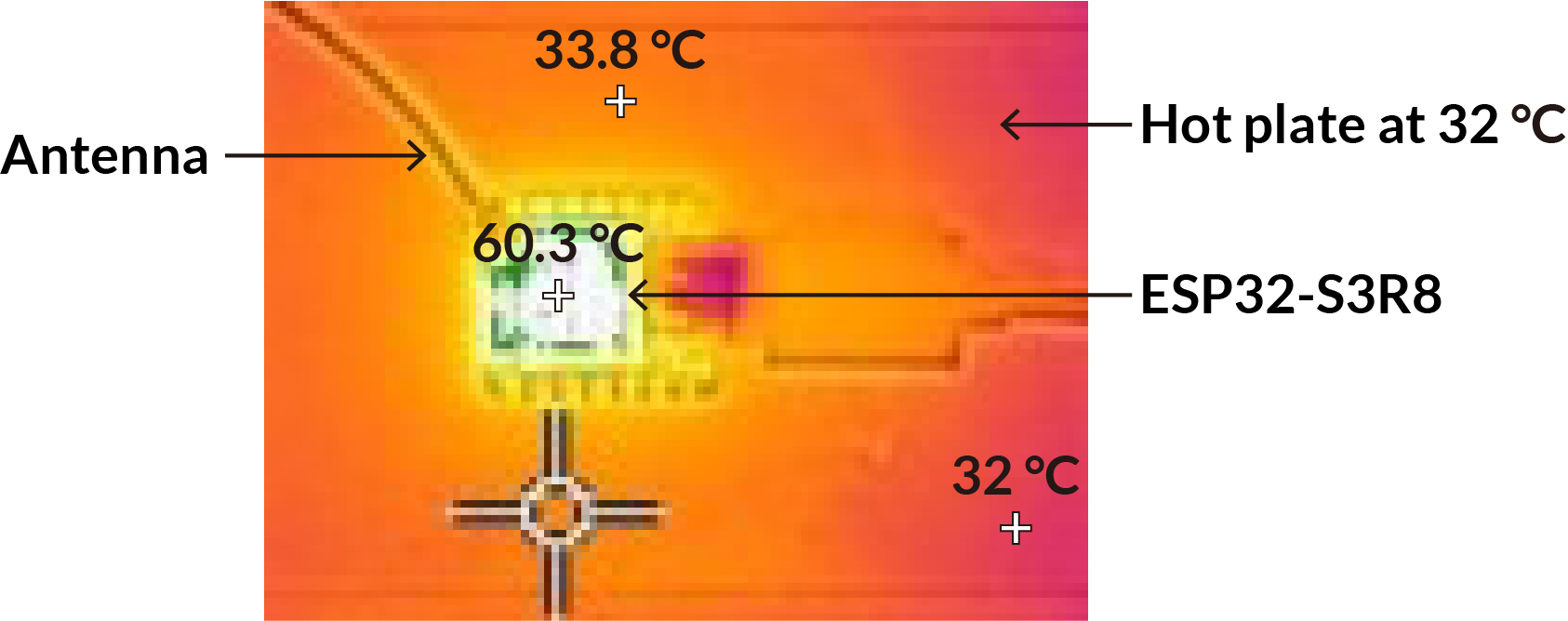}

    \vspace{1em}
        
    \includegraphics[width=0.9\linewidth]{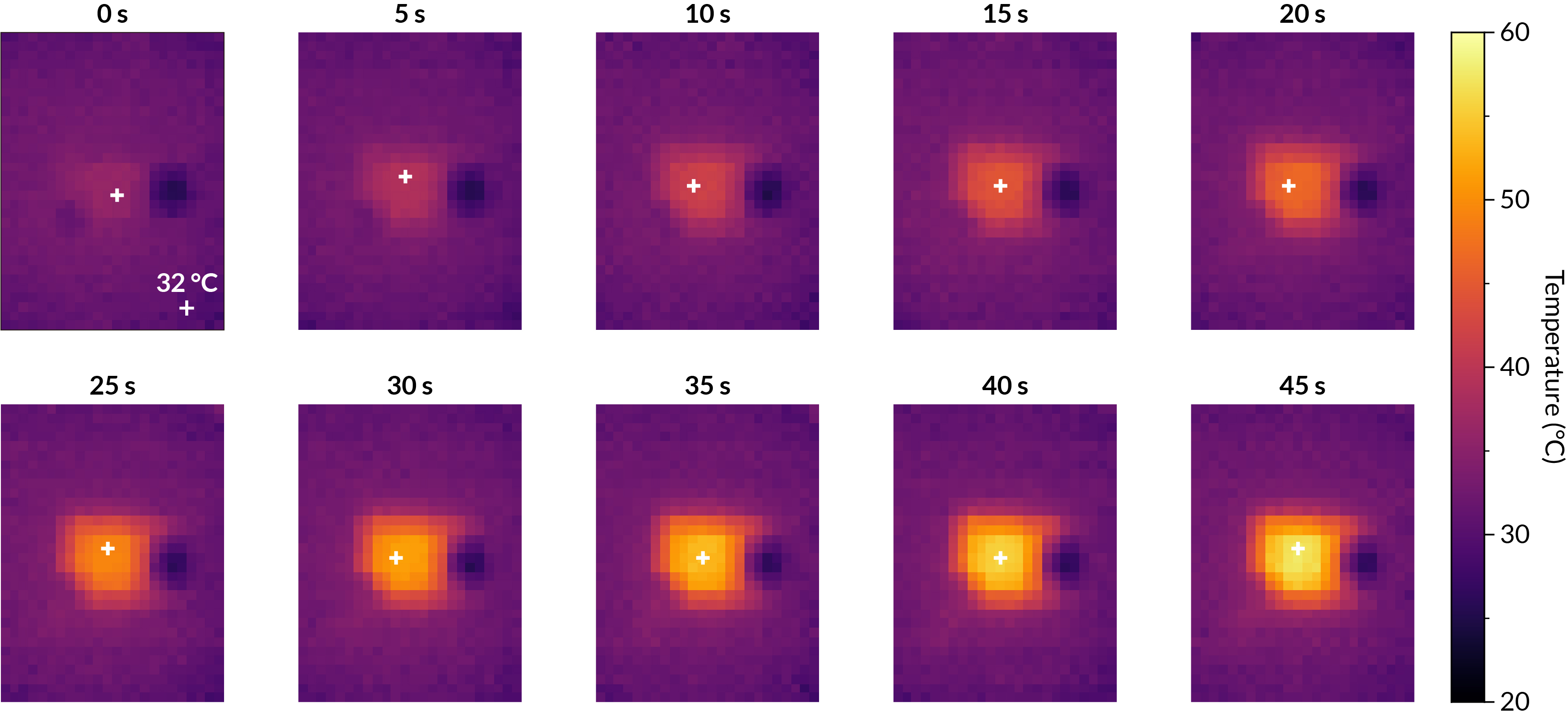}
    \caption{\textbf{Thermal Imaging of the ESP32 Under Contact-Heated Conditions.} Infrared snapshot of the ESP32-S3R8 running the unoptimized TinyLLaMA inference on a $32\,^{\circ}\text{C}$ hot plate, reaching a peak surface temperature of $60.3\,^{\circ}\text{C}$ after 1 min 
    ($\Delta T = +28.3\,^{\circ}\text{C}$).}
    \label{fig:thermal_image_esp32}
\end{figure}

\clearpage
\subsection{More Results}

\begin{table}[h]
    \centering
    \definecolor{customcellcolor}{HTML}{AECBFA}
    \definecolor{highercolor}{HTML}{1967D2}
    \caption{\textbf{Per-Model Deployment Success Rate for Heat Minimization.} Values are successful iterations (\%, ESP32-S3R8); \textcolor{highercolor}{\emph{higher}} is better.}
    \label{tab:heat_success_rate_esp32}
    \begin{subtable}[t]{\columnwidth}
        \centering
        \caption{Room Temp ($T_{\text{amb}} = 25\,^\circ$C), $T_{\text{peak}} \leq 37\,^\circ$C}
        \label{tab:heat_success_esp32_room}
        \begin{tabular}{@{}lcccccc@{}}
        \toprule
        \textbf{Model} & \textbf{Sco} & \textbf{Doc} & \textbf{HIL} \\
        \midrule
        Claude Opus 4.7 & 0 & \cellcolor{customcellcolor!30}30 & \cellcolor{customcellcolor!40}\textbf{40} \\
        Claude Sonnet 4.6 & \cellcolor{customcellcolor!20}20 & 0 & \cellcolor{customcellcolor!20}20 \\
        GPT-5.4 & \cellcolor{customcellcolor!50}50 & \cellcolor{customcellcolor!10}10 & \cellcolor{customcellcolor!70}\textbf{70} \\
        GPT-5.4 mini & 0 & \cellcolor{customcellcolor!20}\textbf{20} & 0 \\
        Gemini 3.1 Pro & \cellcolor{customcellcolor!10}10 & 0 & \cellcolor{customcellcolor!30}\textbf{30} \\
        Gemini 3 Flash & 0 & 0 & 0 \\
        \bottomrule
        \end{tabular}
    \end{subtable}
    
    \vspace{1em}
    
    \begin{subtable}[t]{\columnwidth}
        \centering
        \caption{Contact-heated ($T_{\text{contact}} \approx 33\,^\circ$C), $T_{\text{peak}} \leq 51\,^\circ$C}
        \label{tab:heat_success_esp32_skin}
        \begin{tabular}{@{}lcccccc@{}}
        \toprule
        \textbf{Model} & \textbf{Sco} & \textbf{Doc} & \textbf{HIL} \\
        \midrule
        Claude Opus 4.7 & 0 & 0 & \cellcolor{customcellcolor!50}\textbf{50} \\
        Claude Sonnet 4.6 & 0 & 0 & \cellcolor{customcellcolor!30}\textbf{30} \\
        GPT-5.4 & \cellcolor{customcellcolor!40}40 & \cellcolor{customcellcolor!100}\textbf{100} & \cellcolor{customcellcolor!70}70 \\
        GPT-5.4 mini & 0 & 0 & 0 \\
        Gemini 3.1 Pro & 0 & 0 & \cellcolor{customcellcolor!50}\textbf{50} \\
        Gemini 3 Flash & \cellcolor{customcellcolor!20}\textbf{20} & 0 & 0 \\
        \bottomrule
        \end{tabular}
    \end{subtable}
\end{table}

\begin{table}[h]
    \centering
    \definecolor{customcellcolor}{HTML}{F6AEA9}
    \definecolor{lowercolor}{HTML}{C5221F}
    \caption{\textbf{Best Peak Temperature on ESP32.} Values are $\Delta T = T_{\text{peak}} - T_{\text{amb/contact}}$; \textcolor{lowercolor}{\emph{lower}} is better.}
    \label{tab:heat_best_temp_esp32}
    \begin{subtable}[t]{\columnwidth}
        \centering
        \caption{Room Temp ($T_{\text{amb}} = 25\,^\circ$C), $T_{\text{peak}} \leq 37\,^\circ$C}
        \label{tab:heat_best_temp_esp32_room}
        \begin{tabular}{@{}lcccccc@{}}
        \toprule
        \textbf{Model} & \textbf{Sco} & \textbf{Doc} & \textbf{HIL} \\
        \midrule % (value - 7) * 10
        Claude Opus 4.7 & \redx & \cellcolor{customcellcolor!44}+11.42 (36.42) & \cellcolor{customcellcolor!40}\textbf{+11.04 (36.04)} \\
        Claude Sonnet 4.6 & \cellcolor{customcellcolor!40}+11.00 (36.00) & \redx & \cellcolor{customcellcolor!36}\textbf{+10.59 (35.59)} \\
        GPT-5.4 & \cellcolor{customcellcolor!46}+11.59 (36.59) & \cellcolor{customcellcolor!43}+11.34 (36.34) & \cellcolor{customcellcolor!4}\textbf{+7.41 (32.41)} \\
        GPT-5.4 mini & \redx & \cellcolor{customcellcolor!35}\textbf{+10.48 (35.48)} & \redx \\
        Gemini 3.1 Pro & \cellcolor{customcellcolor!49}+11.85 (36.85) & \redx & \cellcolor{customcellcolor!39}\textbf{+10.86 (35.86)} \\
        Gemini 3 Flash & \redx & \redx & \redx \\
        \bottomrule
        \end{tabular}
    \end{subtable}
    
    \vspace{1em}
    
    \begin{subtable}[t]{\columnwidth}
        \centering
        \caption{Contact-heated ($T_{\text{contact}} \approx 33\,^\circ$C), $T_{\text{peak}} \leq 51\,^\circ$C}
        \label{tab:heat_best_temp_esp32_skin}
        \begin{tabular}{@{}lcccccc@{}}
        \toprule
        \textbf{Model} & \textbf{Sco} & \textbf{Doc} & \textbf{HIL} \\
        \midrule % (value - 11) * 10
        Claude Opus 4.7 & \redx & \redx & \cellcolor{customcellcolor!15}\textbf{+12.46 (44.46)} \\
        Claude Sonnet 4.6 & \redx & \redx & \cellcolor{customcellcolor!74}\textbf{+18.44 (50.44)} \\
        GPT-5.4 & \cellcolor{customcellcolor!51}+16.15 (48.15) & \cellcolor{customcellcolor!11}\textbf{+12.09 (44.09)} & \cellcolor{customcellcolor!54}+16.43 (48.43) \\
        GPT-5.4 mini & \redx & \redx & \redx \\
        Gemini 3.1 Pro & \redx & \redx & \cellcolor{customcellcolor!50}\textbf{+16.03 (48.03)} \\
        Gemini 3 Flash & \cellcolor{customcellcolor!67}\textbf{+17.65 (49.65)} & \redx & \redx \\
        \bottomrule
        \end{tabular}
    \end{subtable}
\end{table}

\begin{table}[h]
    \centering
    \definecolor{customcellcolor}{HTML}{1967D2}
    \caption{\textbf{Per-Model Token Usage for ESP32-S3R8 Heat Minimization.} Values are in millions (M) of tokens.}
    \label{tab:heat_esp32_token}
    \begin{subtable}[t]{\columnwidth}
        \centering
        \caption{Room Temp ($T_{\text{amb}} = 25\,^\circ$C), $T_{\text{peak}} \leq 37\,^\circ$C}
        \label{tab:heat_esp32_token_room}
        \begin{tabular}{@{}lccc@{}}
        \toprule
        \textbf{Model} & \textbf{Sco} & \textbf{Doc} & \textbf{HIL} \\
        \midrule
        Claude Opus 4.7 & \cellcolor{customcellcolor!9}\textbf{8.71} & \cellcolor{customcellcolor!10}9.63 & \cellcolor{customcellcolor!9}9.46 \\
        Claude Sonnet 4.6 & \cellcolor{customcellcolor!4}\textbf{4.14} & \cellcolor{customcellcolor!5}5.25 & \cellcolor{customcellcolor!7}6.74 \\
        GPT-5.4 & \cellcolor{customcellcolor!3}\textbf{2.86} & \cellcolor{customcellcolor!5}5.36 & \cellcolor{customcellcolor!6}5.79 \\
        GPT-5.4 mini & \cellcolor{customcellcolor!6}5.66 & \cellcolor{customcellcolor!4}\textbf{3.63} & \cellcolor{customcellcolor!6}5.82 \\
        Gemini 3.1 Pro & \cellcolor{customcellcolor!11}11.22 & \cellcolor{customcellcolor!10}\textbf{10.16} & \cellcolor{customcellcolor!15}14.70 \\
        Gemini 3 Flash & \cellcolor{customcellcolor!10}10.06 & \cellcolor{customcellcolor!8}\textbf{7.91} & \cellcolor{customcellcolor!11}11.23 \\
        \bottomrule
        \end{tabular}
    \end{subtable}
    
    \vspace{1em}
    
    \begin{subtable}[t]{\columnwidth}
        \centering
        \caption{Contact-heated ($T_{\text{contact}} \approx 33\,^\circ$C), $T_{\text{peak}} \leq 51\,^\circ$C}
        \label{tab:heat_esp32_token_skin}
        \begin{tabular}{@{}lccc@{}}
        \toprule
        \textbf{Model} & \textbf{Sco} & \textbf{Doc} & \textbf{HIL} \\
        \midrule
        Claude Opus 4.7 & \cellcolor{customcellcolor!7}\textbf{7.34} & \cellcolor{customcellcolor!11}11.11 & \cellcolor{customcellcolor!12}12.22 \\
        Claude Sonnet 4.6 & \cellcolor{customcellcolor!8}7.50 & \cellcolor{customcellcolor!7}7.13 & \cellcolor{customcellcolor!6}\textbf{5.78} \\
        GPT-5.4 & \cellcolor{customcellcolor!6}\textbf{5.52} & \cellcolor{customcellcolor!6}5.77 & \cellcolor{customcellcolor!8}7.62 \\
        GPT-5.4 mini & \cellcolor{customcellcolor!6}5.75 & \cellcolor{customcellcolor!4}\textbf{4.49} & \cellcolor{customcellcolor!8}7.88 \\
        Gemini 3.1 Pro & \cellcolor{customcellcolor!11}11.00 & \cellcolor{customcellcolor!10}\textbf{10.49} & \cellcolor{customcellcolor!14}13.62 \\
        Gemini 3 Flash & \cellcolor{customcellcolor!14}13.89 & \cellcolor{customcellcolor!10}\textbf{10.02} & \cellcolor{customcellcolor!11}11.32 \\
        \bottomrule
        \end{tabular}
    \end{subtable}
\end{table}

\begin{table}[h]
    \centering
    \definecolor{customcellcolor}{HTML}{1967D2}
    \caption{\textbf{Per-Model Tool Use for ESP32-S3R8 Heat Minimization.} Values represent the average number of tool calls per iteration.}
    \label{tab:esp32_token}
    \begin{subtable}[t]{\columnwidth}
        \centering
        \caption{Room Temp ($T_{\text{amb}} = 25\,^\circ$C), $T_{\text{peak}} \leq 37\,^\circ$C}
        \label{tab:esp_toolcall_room}
        \begin{tabular}{@{}lccc@{}}
        \toprule
        \textbf{Model} & \textbf{Sco} & \textbf{Doc} & \textbf{HIL} \\
        \midrule
        Claude Opus 4.7 & \cellcolor{customcellcolor!11}\textbf{11.1} & \cellcolor{customcellcolor!14}14.1 & \cellcolor{customcellcolor!12}11.7 \\
        Claude Sonnet 4.6 & \cellcolor{customcellcolor!7}\textbf{7.0} & \cellcolor{customcellcolor!9}9.1 & \cellcolor{customcellcolor!12}11.7 \\
        GPT-5.4 & \cellcolor{customcellcolor!7}\textbf{7.4} & \cellcolor{customcellcolor!13}12.5 & \cellcolor{customcellcolor!12}12.4 \\
        GPT-5.4 mini & \cellcolor{customcellcolor!18}17.8 & \cellcolor{customcellcolor!12}11.7 & \cellcolor{customcellcolor!11}\textbf{10.9} \\
        Gemini 3.1 Pro & \cellcolor{customcellcolor!20}20.0 & \cellcolor{customcellcolor!16}\textbf{16.3} & \cellcolor{customcellcolor!18}18.0 \\
        Gemini 3 Flash & \cellcolor{customcellcolor!15}15.4 & \cellcolor{customcellcolor!14}13.5 & \cellcolor{customcellcolor!12}\textbf{12.3} \\
        \bottomrule
        \end{tabular}
    \end{subtable}
    
    \vspace{1em}
    
    \begin{subtable}[t]{\columnwidth}
        \centering
        \caption{Contact-heated ($T_{\text{contact}} \approx 33\,^\circ$C), $T_{\text{peak}} \leq 51\,^\circ$C}
        \label{tab:esp_toolcall_skin}
        \begin{tabular}{@{}lccc@{}}
        \toprule
        \textbf{Model} & \textbf{Sco} & \textbf{Doc} & \textbf{HIL} \\
        \midrule
        Claude Opus 4.7 & \cellcolor{customcellcolor!12}\textbf{12.2} & \cellcolor{customcellcolor!15}14.8 & \cellcolor{customcellcolor!14}14.0 \\
        Claude Sonnet 4.6 & \cellcolor{customcellcolor!16}15.6 & \cellcolor{customcellcolor!15}14.6 & \cellcolor{customcellcolor!10}\textbf{9.9} \\
        GPT-5.4 & \cellcolor{customcellcolor!14}\textbf{14.2} & \cellcolor{customcellcolor!16}15.7 & \cellcolor{customcellcolor!16}16.0 \\
        GPT-5.4 mini & \cellcolor{customcellcolor!14}\textbf{13.9} & \cellcolor{customcellcolor!15}14.8 & \cellcolor{customcellcolor!20}19.5 \\
        Gemini 3.1 Pro & \cellcolor{customcellcolor!20}20.0 & \cellcolor{customcellcolor!20}20.1 & \cellcolor{customcellcolor!19}\textbf{19.0} \\
        Gemini 3 Flash & \cellcolor{customcellcolor!20}19.9 & \cellcolor{customcellcolor!15}14.8 & \cellcolor{customcellcolor!14}\textbf{13.9} \\
        \bottomrule
        \end{tabular}
    \end{subtable}
\end{table}

\clearpage
\subsection{Failure Modes in Thermal Management}

Because thermal dissipation is intrinsically linked to the underlying power profile ($Q = \int_{0}^{\tau} P(t)\,\mathrm{d}t$), the thermal optimization task on the ESP32-S3R8 shares the fundamental failure modes detailed in Power Minimization (Appendix~\ref{app:power_details}). However, the rigid temporal constraints introduce a task-specific manifestation of these failures:

\begin{itemize}
    \item \textbf{Temporal Deadline Violations:} To mitigate heat generation, agents employ valid thermal management strategies, such as throttling CPU/APB clocks or injecting task delays. However, they sometimes struggle to balance these thermal reductions against the rigid execution constraints. While these trials successfully lower peak temperature ($\Delta T$), they can cause the complex LLM inference and UDP streaming workload to violate the strict 60\,s execution window, resulting in a failed evaluation. Crucially, agents do not always recognize that the failure is due to a latency violation rather than a firmware error; instead of reverting, they may persist down this invalid optimization track and exhaust their iteration budget.
\end{itemize}

\clearpage
\section{Structured Natural Language Feedback Examples}
\label{sec:example_feedback_observation}

After each iteration, the benchmark appends a structured natural-language observation $o_t$ to the agent context. The exact fields depend on the active checks, but the observation always reports which checks ran, whether they passed, the scalar optimization score, and any diagnostic feedback enabled for that experiment. The feedback is fully automated and does not depend on human intervention. Iteration state is formatted as follows with $o_t$ represented by the \texttt{last\_iteration.checks} field:
\begingroup\scriptsize
\begin{verbatim}
Iteration status summary:
{
  "iteration": 1,
  "prior_iterations": [ ... ],
  "last_iteration": {
    "iteration": 0,
    "inputs": { ... },
    "notes": "...",
    "checks": { ... },
  },
  "best_scores_by_check": { ... },
  "current_sandbox_files": [ ... ],
  "tool_budget": { ... },
  "repeated_submission_warning": null,
  "next_action_requirements": [ ... ]
}
\end{verbatim}
\endgroup
The \texttt{prior\_iterations} field is a summary of the prior $(s, o)$ pairs so the agent has a memory of the multi-turn interaction. To help the agent, we auto-compute the best scores from the memory in \texttt{best\_scores\_by\_check} and summarize the available files in the sandbox in \texttt{current\_sandbox\_files}. We also track the remaining tool uses, \texttt{tool\_budget}, and set a warning flag if the agent ever submits identical artifacts via \texttt{repeated\_submission\_warning}. The texttt{next\_action\_requirements} field is an automatically generated enumeration of the failing checks that need to pass. \texttt{last\_iteration.inputs} and the agent generated notes for itself, \texttt{last\_iteration.notes}, constitute the most recently submitted artifact $s_t$. Below we expand the \texttt{last\_iteration.checks} field, $o_t$, for the MAX78000-compression and min-peak-power tasks and three information settings. Ellipses remove long paths, compiler output, or UART lines for presentation only.

\newpage
\subsection{MAX78000 Model Compression}

\paragraph{Score-only (Sco).} Sco in the MAX78000 model compression setting similarly only includes pass/fail and normalized scores, but for different checks: 1) whether the model has trainable gradients, 2) how well the trained model scores on the held-out test set, and 3) whether the model can run on device.
\begingroup\scriptsize
\begin{verbatim}
"checks": {
  "gradient_flow.py": {
    "success": true,
    "score": 1.0,
    "score_unit": "fraction",
    "feedback": null,
    "feedback_image_paths": []
  },
  "train.py": {
    "success": true,
    "score": 0.01484375,
    "score_unit": "mAP50-95",
    "feedback": null,
    "feedback_image_paths": []
  },
  "synthesis_max78000.py": {
    "success": false,
    "score": 0.6969696969696969,
    "score_unit": "fraction",
    "feedback": null,
    "feedback_image_paths": []
  }
}
\end{verbatim}
\endgroup
The Doc $o_t$ is identical since it only changes the initial artifact $s_0$ to include static documentation.

\paragraph{Hardware-in-the-loop (HIL).} HIL exposes each of the sub-steps within each check and reveals whether they were successful. Extra statistics such as the loss curves and GPU utilization are also surfaced to the model so it can tune the training parameters.
\begingroup\scriptsize
\begin{verbatim}
"checks": {
  "gradient_flow.py": {
    "success": true,
    "score": 1.0,
    "score_unit": "fraction",
    "feedback":
      "[pass] model file exists: .../model.py\n"
      "[pass] model file compiles: Python file compiles\n"
      "[pass] gradient smoke test: "
      "trainable_parameters=3928 "
      "input_shape=(2, 3, 32, 32)\n",
    "feedback_image_paths": []
  },
  "train.py": {
    "success": true,
    "score": 0.01484375,
    "score_unit": "mAP50-95",
    "feedback":
      "backend=remote gpu=1 dataset=coco train_examples=512 "
      "test_examples=512 epochs=5 metric=mAP50-95 "
      "initial_score=0.0000 final_score=0.0148 "
      "torch_device=cuda gpu_peak_memory_mb=548 "
      "gpu_peak_utilization_percent=17 "
      "gpu_peak_power_watts=71.4 gpu_telemetry_samples=11",
    "feedback_image_paths": []
  },
  "synthesis_max78000.py": {
    "success": false,
    "score": 0.6969696969696969,
    "score_unit": "fraction",
    "feedback":
      "[pass] model file exists: .../model.py\n"
      "[pass] model file compiles when applicable: "
      "Python file compiles\n"
      "[pass] MAX78000 YAML exists: .../model.yaml\n"
      "[pass] MAX78000 YAML parses: valid YAML\n"
      "[pass] YAML contains architecture fields: "
      "found architecture-like keys\n"
      "[pass] sample input exists: .../sample_input.npy\n"
      "[pass] generated ai8x checkpoint from Python model: "
      ".../.generated/candidate.pth.tar\n"
      "[fail] ai8xize synthesis: ...",
    "feedback_image_paths": []
  }
}
\end{verbatim}
\endgroup

\newpage
\subsection{Peak Current Minimization}
\paragraph{Score-only (Sco).} Sco exposes pass/fail and the normalized scalar score, but not the measured current, UART trace, or compiler/flasher diagnostics.
\begingroup\scriptsize
\begin{verbatim}
"checks": {
  "compile_max78000.py": {
    "success": true,
    "score": 1.0,
    "score_unit": "binary",
    "feedback": null,
    "feedback_image_paths": []
  },
  "flash_max78000.py": {
    "success": true,
    "score": 1.0,
    "score_unit": "binary",
    "feedback": null,
    "feedback_image_paths": []
  },
  "measure_max78000.py": {
    "success": true,
    "score": 0.98237,
    "score_unit": "metric_score(peak_uA)",
    "feedback": null,
    "feedback_image_paths": []
  }
}
\end{verbatim}
\endgroup
The Doc $o_t$ is identical since it only changes the initial artifact $s_0$ to include static documentation.

\paragraph{Hardware-in-the-loop (HIL).}
HIL changes the measurement check to include the raw current and UART Serial Output for debugging/profiling. 
\begingroup\scriptsize
\begin{verbatim}
"measure_max78000.py": {
  "success": true,
  "score": 0.98237,
  "score_unit": "metric_score(peak_uA)",
  "feedback":
    "## Hardware Status\n"
    "capture_power=True, capture_serial=True, "
    "capture_thermal=False\n"
    "duration_ms=20000, optimization_metric=peak_uA, "
    "metric_value=21254.9\n\n"
    "## Measurement Results\n"
    "{\n"
    "  \"avg_uA\": 6254.12,\n"
    "  \"peak_uA\": 21254.87,\n"
    "  \"total_energy_j\": 0.4128,\n"
    "  \"power_samples\": 2000,\n"
    "  \"ppk_sample_rate_hz\": 100.0\n"
    "}\n\n"
    "## Firmware Behavior\n"
    "Checkpoint: FOUND\n"
    "Judge: PASS: UART is consistent with five "
    "live camera inference cycles and "
    "post-processing.\n\n"
    "## UART Serial Output\n"
    "[0000.000] boot\n"
    "[0002.001] camera init complete\n"
    "...\n"
    "[0014.935] firmware task complete checkpoint\n"
    "[0015.036] entering final sleep",
  "feedback_image_paths": []
}
\end{verbatim}
\endgroup
The compile and flash checks only change to include the error message, if any, in the respective \texttt{feedback} subfields.

\clearpage
\section{MAX78000 MCU}
The MAX78000 is a commercially available MCU with an integrated neural accelerator that has been widely adopted by the embedded systems community.

\subsection{Model Size}
\begin{figure}[h]
  \centering
  \includegraphics[width=\linewidth]{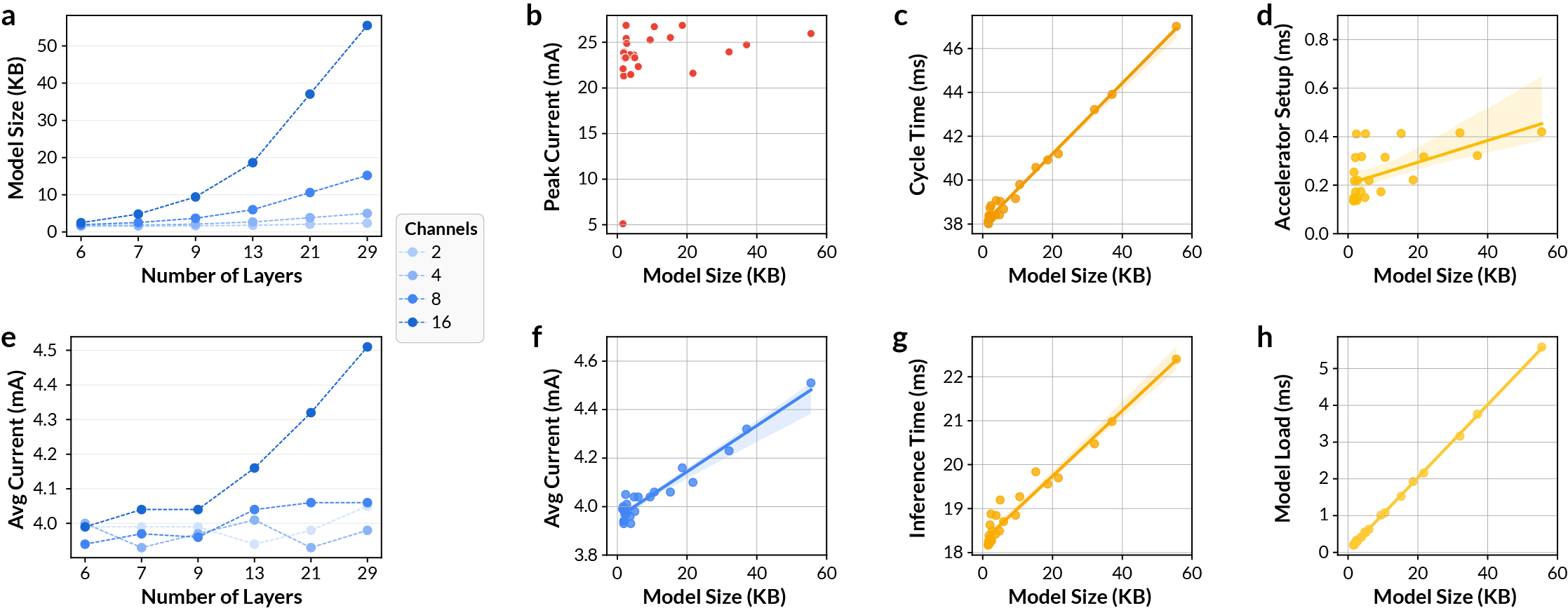}
  \caption{\textbf{Impact of model architecture on size and physical performance.}
  (a) Model size versus network depth for different channel widths, demonstrating that size scales with depth.
  (b) Peak current consumption shows minimal correlation with model size.
  (c-h) Total cycle time (c), comprising accelerator setup (d), model weight loading (h), and inference (g), as well as average current consumption during inference, all scale linearly with model size (N=10).
  }
  \label{fig:power_vs_model_size}
\end{figure}

We show how model depths (i.e., the number of layers) and channel dimensions affect model size and runtime characteristics on the MCU (Fig.~\ref{fig:power_vs_model_size}).
Contrary to intuition, peak current consumption shows minimal correlation with model size (Fig.~\ref{fig:power_vs_model_size}b), remaining consistently below 30 mA regardless of size.
Figure~\ref{fig:power_vs_model_size}c-h show that total inference time and power consumption scale approximately linearly with model size.

\clearpage
\subsection{Clock Frequency}
A critical configurable parameter of the MAX78000 is its clock source, which determines both the CNN accelerator's execution speed and power profile. The MCU supports two primary clock routing modes for the neural hardware:
\begin{itemize}
    \item \textbf{System Clock Source:} The accelerator clock is derived from the main system clock via a hardware $\div 2$ divider. For instance, configuring a system clock of 60 MHz yields an effective CNN clock of 30 MHz.
    \item \textbf{Direct 60 MHz Source:} A dedicated 60 MHz oscillator is routed directly to the accelerator, bypassing the frequency divider. This is denoted as "60 (Dir)" in our evaluations.
\end{itemize}

\begin{figure*}[h]
  \centering
  \includegraphics[width=0.7\linewidth]{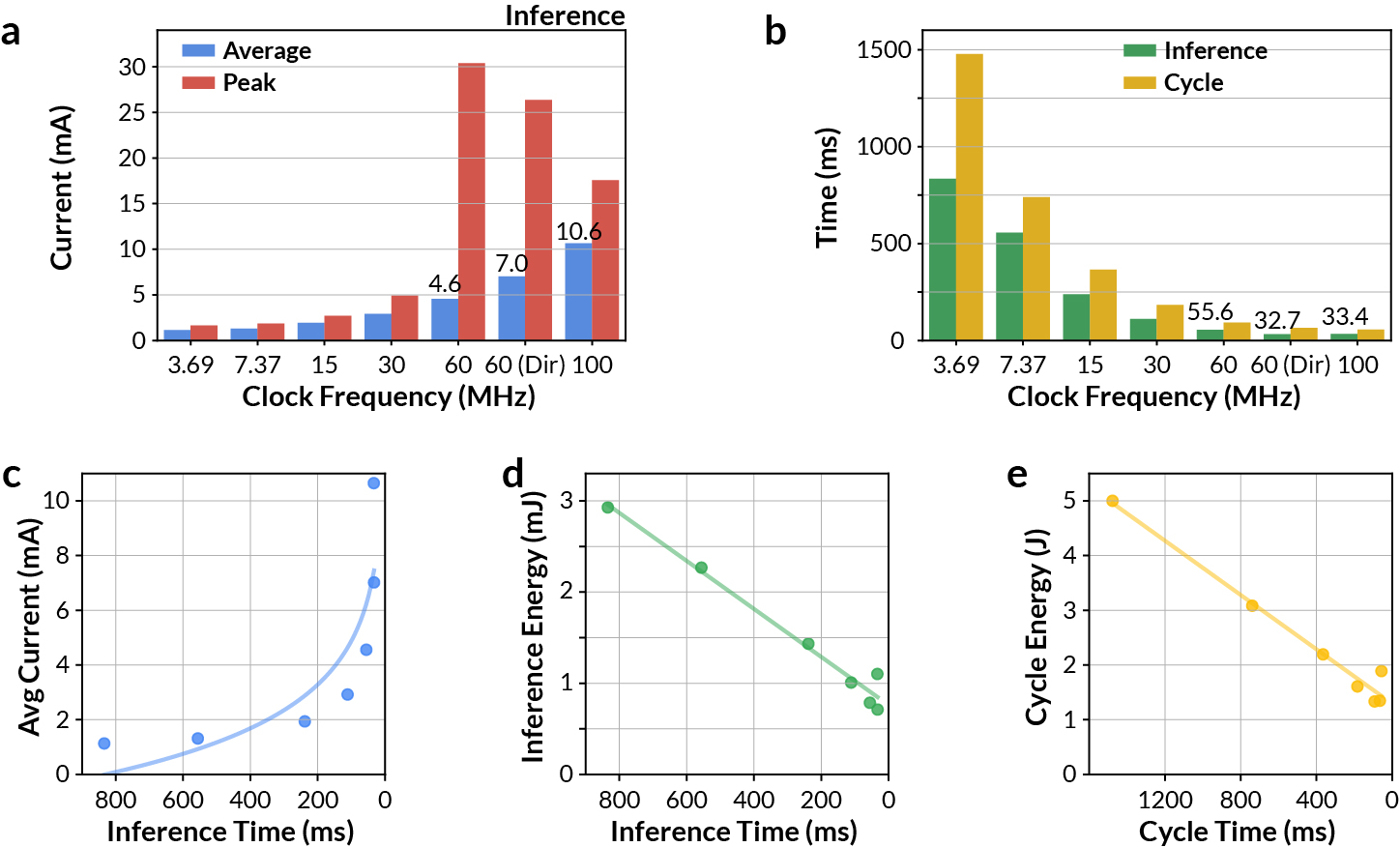}
  \caption{Impact of system clock frequency on model inference performance and energy consumption. 
  (a, b) Average and peak current consumption (a) and inference and cycle time (b) at different system clock frequencies (N=10). 
  Higher clock frequencies generally increase instantaneous current but decrease completion time. 
  (c-e) Plots of average current (c), inference energy (d), and cycle energy (e) versus time demonstrate that faster completion times result in lower total energy consumption, despite higher instantaneous current. 
  "60 (Dir)" denotes a 60 MHz clock set as the system clock and also applied directly to the CNN accelerator without frequency division. All other labels (including "60") represent the set system clock frequency, which is then divided by two to produce the final CNN clock. All measurements were performed on the same YOLO detection model.
  }
  \label{fig:inference_power_vs_clock_freq}
\end{figure*}

As shown in Fig.~\ref{fig:inference_power_vs_clock_freq}, the selected clocking scheme creates a counterintuitive trade-off between instantaneous power and total energy. Operating the accelerator at higher frequencies (e.g., the direct 60 MHz source) predictably increases the peak and average instantaneous current. However, this is heavily offset by a dramatic reduction in computation latency. Because the inference cycle completes significantly faster, the total energy consumed per inference is actually minimized at higher clock speeds. This dynamic is a critical consideration for battery-free deployments: systems should ideally run at the maximum possible clock frequency to minimize total energy, provided the energy harvester and power management circuitry can sustain the higher instantaneous peak current demands.

\clearpage
\section{Wildlife Monitoring: Elk Detection}\label{app:elk}
Camera traps are indispensable tools in ecology, wildlife research, and conservation, enabling non-invasive observation of animal behavior. However, deployment at scale is limited by cost, battery life, and maintenance requirements that become prohibitive in large or difficult-to-access locations~\cite{glover-kapfer_camera-trapping_2019}.

\emph{We engaged with a commercial farmer in the Pacific Northwest} managing multiple agricultural fields totaling thousands of acres, where elk frequently damage crops. The farmer had deployed five traditional camera trap units at a capital cost of ~\$1,000 each, but three pain points limited scaling:

\begin{enumerate}
    \item \textbf{Energy and Maintenance.} Traditional trail cameras require battery replacement every month under active use. For the farmer's deployment scenario with cameras distributed across large, remote fields, this represents substantial labor overhead. While some cameras near field edges could be connected to power via wires, this is infeasible for most monitoring locations far from electrical infrastructure.
    \item \textbf{Communication Costs.} To enable real-time monitoring without physical visits, cellular-connected cameras transmit all captured images to cloud storage, currently incurring approximately \$10 per camera per month. The farmer estimated that camera coverage every 200 feet would be needed to reliably detect elk near trails along field perimeters. For a single 200-acre field, this translates to 50 cameras and over \$500 monthly communication costs, economically unsustainable when scaled across multiple fields.
    \item \textbf{False Triggers.} The cameras produce large volumes of images containing no wildlife of interest, triggered by wind-blown vegetation or non-target animals (e.g., deer, birds). These false positives consume storage, bandwidth, and review time.
\end{enumerate}

Battery-free model on MCUs directly addresses all three:
1) solar harvesting eliminates battery replacement,
2) local classification reduces cloud transmission for most images, and
3) accurate on-device inference filters false triggers before transmission.

We deployed our agentic pipeline to compress a YOLOv8 binary classification model trained on a curated elk dataset, achieving 96.7\% accuracy within the MAX78000's memory and power constraints. 
Deployment challenges remain: the farmer's traditional camera units failed after a few months when moisture penetrated the enclosure, and developing weatherproof enclosures for our hardware is ongoing. Long-term reliability and return on investment relative to current practices require field validation, which will be reported in follow-up work.

\clearpage
\subsection{Custom MCU Platform} \label{sec:max78000_pcb_details}

\begin{figure}[h]           
  \centering
  \includegraphics[width=\linewidth]{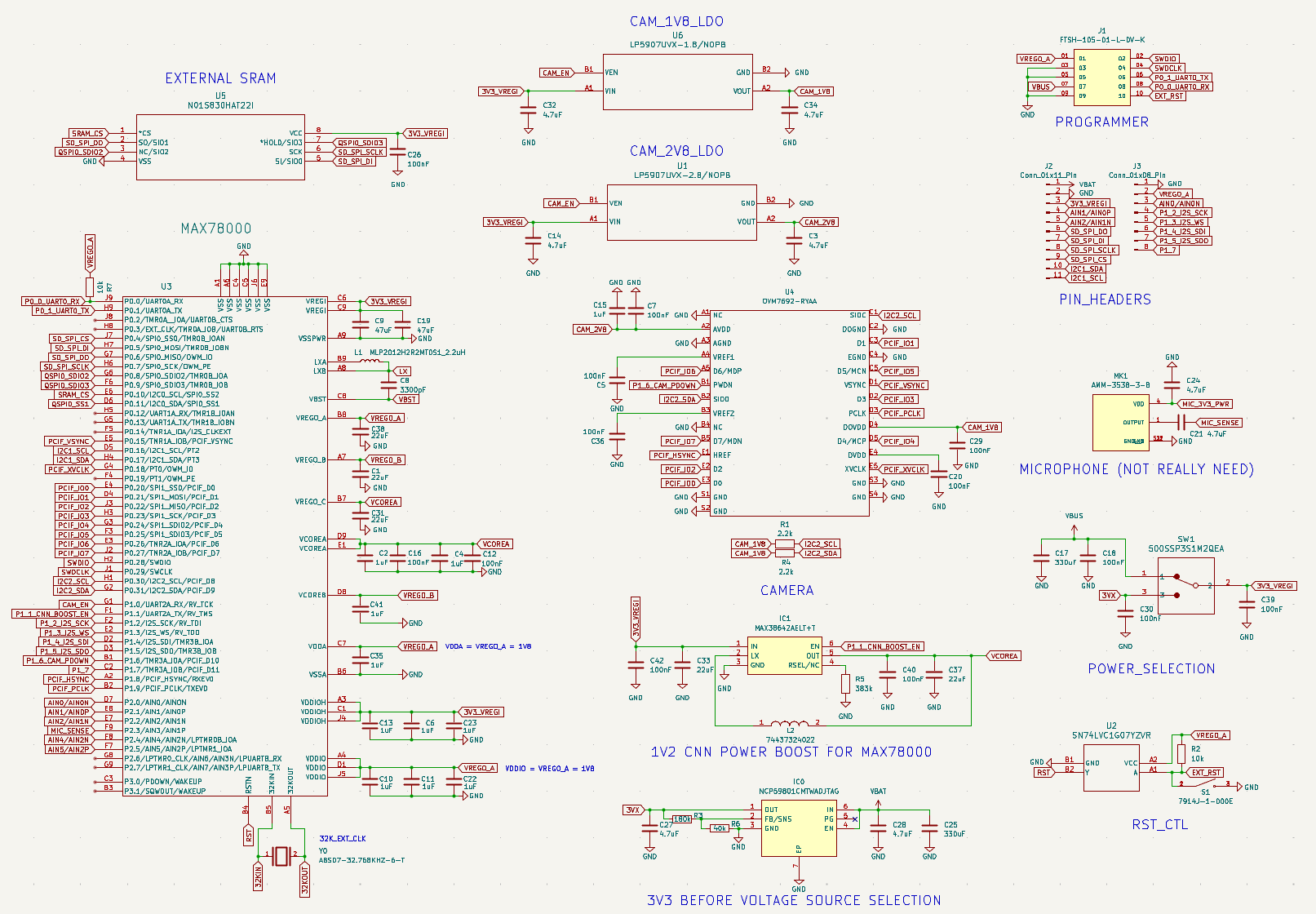}
  \caption{Schematic Design. The schematic comprises two primary blocks: a power module and a functional module. The functional module integrates the MAX78000 SoC, the OVM7692\_RYAA camera, external SRAM, and a SWD interface for programming and debugging.}
  \label{fig:max78000_pcb_schematic}
\end{figure}

\begin{figure}[h]            
  \centering
  \includegraphics[width=0.4\linewidth]{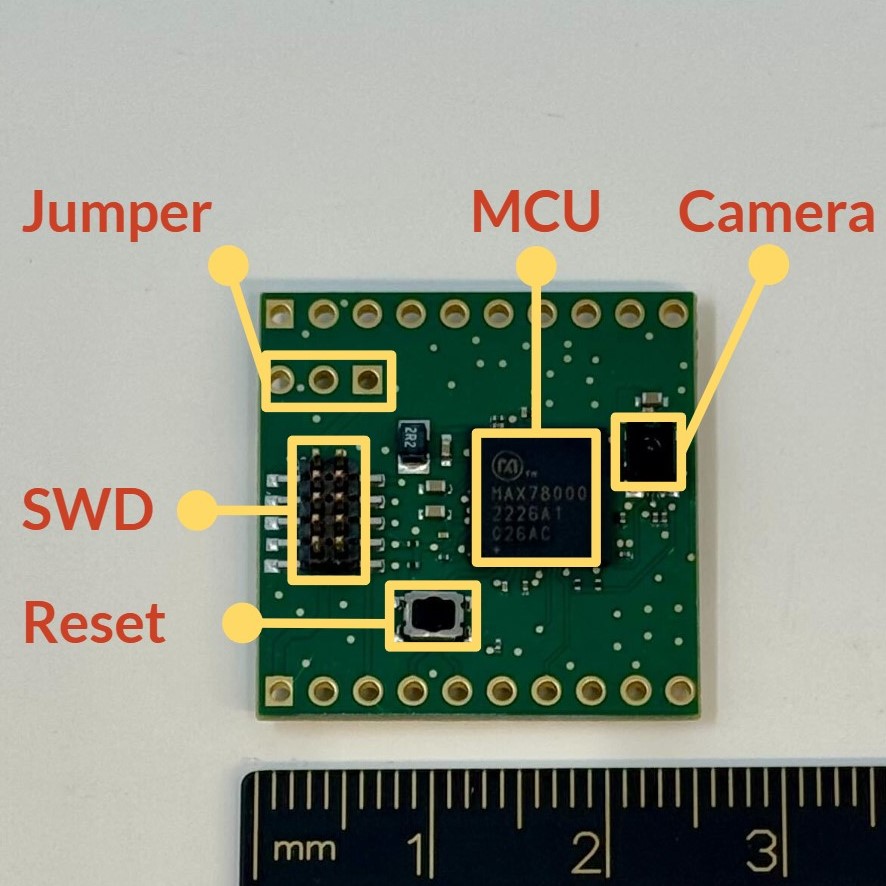} 
  \caption{MAX78000 PCB Platform for wildlife monitoring.}
  \label{fig:max78000_pcb_labels}
\end{figure}

\clearpage
\subsection{Camera}
The GC0308 camera module is interfaced with the MAX78000 via the Parallel Camera Interface (PCIF). The camera's power consumption and capture time depend on three key parameters: image resolution, PCIF frequency, and MCU system clock frequency. Frame rate can be set via duty cycle based on the application. 

\begin{figure}[h]
  \centering
  \includegraphics[width=\linewidth]{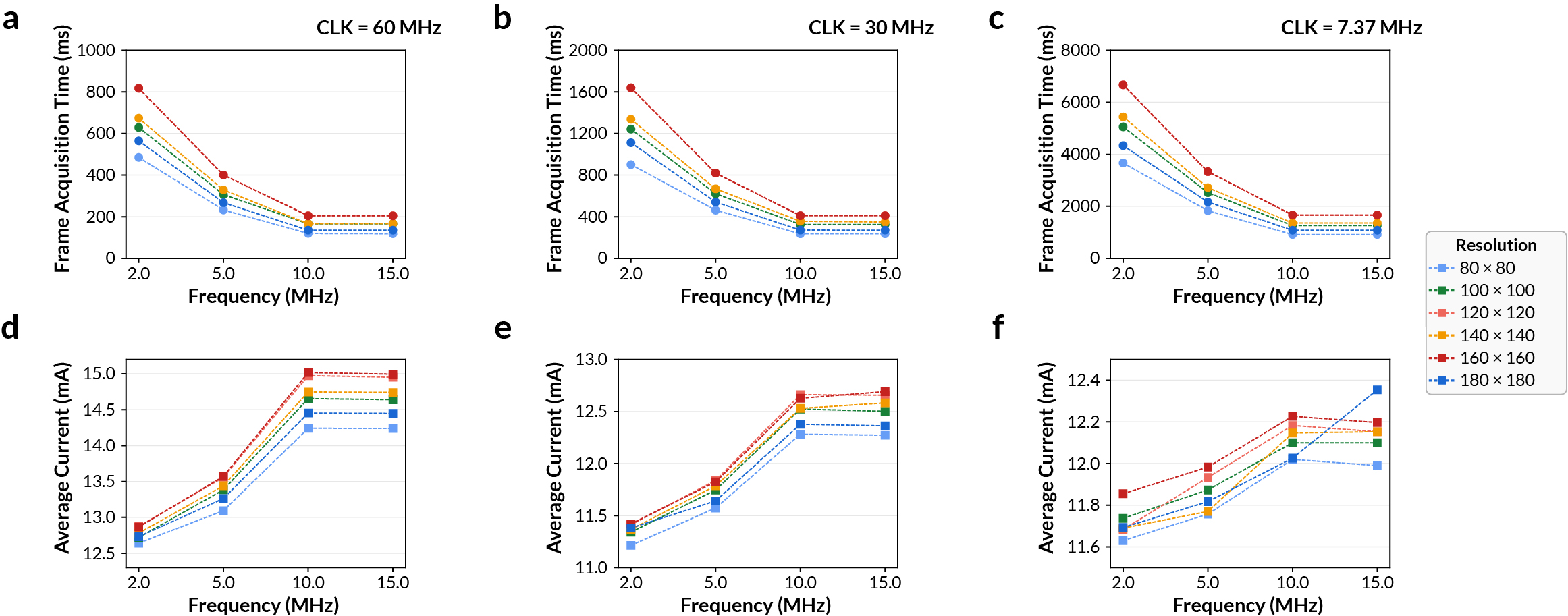}
  \caption{\textbf{Camera acquisition performance on MAX78000 across system clocks, PCIF frequencies, and image resolutions.} (a-c) Frame acquisition time (including camera capture and wake-up overhead) versus PCIF frequency. (d-f) Average current consumption per frame (N=10).
  Resolutions are hardware-downsampled from the camera's native output, demonstrating that metrics mainly depend on the native sensor resolution rather than the final downsampled output.
  }
  \vspace{-0.3cm}
  \label{fig:camera_freq}
\end{figure}

Resolution, however, affects both camera power consumption and vision model accuracy. Figure~\ref{fig:camera_freq}a-c shows frame acquisition time across different resolutions and PCIF frequencies. Notably, all resolutions we evaluated use a hardware-built-in downsampling from the camera's native output, which is an unavoidable feature given by the MSDK driver.
The acquisition time is determined primarily by the native sensor resolution rather than the downsampled output size. For example, a 100×100 image downsampled from 400×400 takes longer than a 180×180 image downsampled from 360×360, because the camera sensor must still read and process the full array before downsampling.
Average current consumption per frame (Fig.~\ref{fig:camera_freq}d-f) follows similar patterns. 

Lower native resolutions reduce both capture time and total power.

\paragraph{Clock Frequency.}
Higher PCIF frequencies reduce frame transfer time from the camera to the MCU memory. With 60 MHz MCU system clock, at 10 MHz PCIF, a 180×180 frame requires ~135.5 ms to capture and transfer, while at 5 MHz PCIF this doubles to 267.7 ms (Fig.~\ref{fig:camera_freq}a). However, higher PCIF frequencies slightly increase instantaneous current draw by 1.19 mA due to increased switching activity in the PCIF peripheral and I/O buffers.

Compared to the optimal 60 MHz system clock we found, reducing to 30 MHz decreases average current by 14.4\% but doubles the capture and total cycle time, leading to higher overall energy consumption.
At 7.37 MHz, cycle time balloons to over 1 second, a 7× increase, making real-time applications infeasible despite slightly lower instantaneous power.
We note that the system clock affects DMA throughput and memory access latency during frame buffer transfers. Lower clocks create memory access bottlenecks that dramatically extend capture time, overwhelming any current reduction benefits.

\clearpage
\subsection{LoRa Communication} \label{subsec:lora_details}
An SX1279 LoRa transceiver operated at 3.3 V is interfaced with the MAX78000 MCU via SPI.

\begin{figure*}[h]
    \centering
    \includegraphics[width=0.5\linewidth]{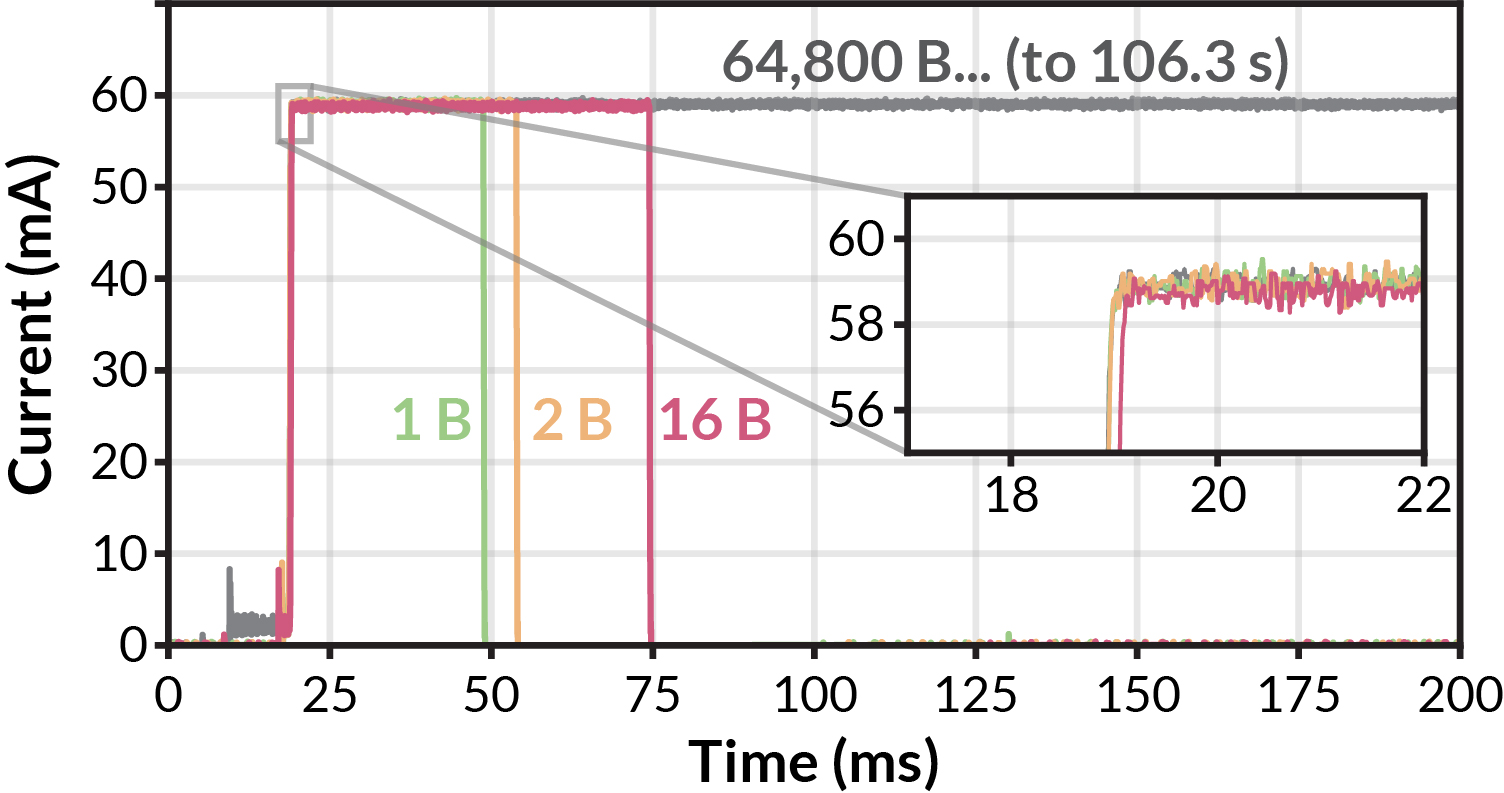}
    
    \vspace{1em}
 
    \begin{tabular}{@{}l|l|l|l@{}}
    \toprule
    \textbf{Packet size} & \textbf{Tx time (ms)} & \textbf{$I_{\text{tx,avg}}$ (mA)} & \textbf{$E_{\text{tx}}$ (mJ)} \\
    \midrule
    \rowcolor[HTML]{F3F3F3}
    1 B & $30.122 \pm 0.013$ & $58.968 \pm 0.027$ & 5.862 \\
    \rowcolor[HTML]{FFFFFF}
    2 B & $35.500 \pm 0.012$ & $59.148 \pm 0.025$ & 6.929 \\
    \rowcolor[HTML]{F3F3F3}
    16 B & $56.836 \pm 0.011$ & $59.302 \pm 0.026$ & 11.123 \\
    \rowcolor[HTML]{FFFFFF}
    64{,}800 B & $106{,}321.2 \pm 0.837$ & $57.656 \pm 0.027$ & 20229.182 \\
    \bottomrule
    \end{tabular}
    \caption{Current consumption profile during LoRA transmission for payload sizes of 1 B, 2 B, 16 B, and 64,800 B (N=10). The 64{,}800 B case represents a large data transfer segmented into multiple LoRA packets rather than a single-packet transmission. Peak transmission current remains below 60 mA across all payload sizes. Small payloads complete within 60 ms, while the 64,800 B transmission continues to 106.3 s (axis truncated for visualization).}
    \label{fig:lora_transmit}
\end{figure*}

\subsubsection{SPI Interface Configuration and Debugging}
The SPI connection was configured in Mode 0 at 1 MHz. A minimal custom driver was developed because initial integration attempts using the RadioLib library failed to establish communication. 
For connectivity verification, RegVersion (address 0x42; datasheet default 0x12) was read, but the initial readout deviated from expectation. 
Oscilloscope measurements of the clock (SCLK), chip select (CS), master output (MOSI), and master input (MISO) signals confirmed correct timing and frame formatting from the microcontroller while capturing the transceiver responses.
Analysis revealed that the first byte returned after power-up or upon entering LoRa mode, the first byte returned by the first read transaction was consistently invalid, whereas all subsequent bytes were correct (0x12 for RegVersion). 
To address this hardware quirk, a warm-up procedure was added to the initialization sequence: a dummy read of RegVersion is performed immediately after reset or mode switching, with the first returned byte discarded before normal register operations begin. 
With this step in place, the register I/O and the TX/RX data path operated reliably and reproducibly.

\subsubsection{LoRa Driver Architecture}
The custom driver implemented four primary operational phases:

\paragraph{Initialization.} Following a hardware reset, the transceiver entered LoRa Sleep mode, then transitioned to Standby. A warm-up dummy SPI read was performed with the first byte discarded.

\paragraph{Configuration.} The radio was configured with a center frequency of 915\,MHz. Modem parameters were set as follows: bandwidth (BW) = 125\,kHz, coding rate (CR) = 4/5, spreading factor (SF) = 7, explicit header mode enabled, cyclic redundancy check (CRC) enabled, and preamble length = 12 symbols.

\paragraph{Transmission.} From Standby state, the interrupt flags were cleared, the FIFO pointer was configured, payload data and length were written to the buffer, and the radio transitioned to TX mode. The system awaited the TxDone interrupt (with timeout protection), cleared status flags, and returned to Standby or Sleep mode.

\paragraph{Reception.} Reception follows a similar state machine: flags are cleared, the receive pointer is reset, continuous or single receive mode is entered, the system waits for reception completion with timeout, payload integrity is verified via cyclic redundancy check, data is read from the FIFO buffer, flags are cleared, and the radio optionally enters Sleep mode.

\subsubsection{Power Minimization for LoRa Driver}
To balance energy consumption with long-range communication reliability, SF7 with BW = 125\,kHz, CR = 4/5, and 12-symbol preamble was selected. 
At 125\,kHz bandwidth, SF7 yielded a symbol duration of 1.024\,ms, reducing time-on-air and per-message energy consumption compared to higher spreading factors. The CR = 4/5 configuration provided the minimum forward error correction overhead while maintaining robustness against typical outdoor channel impairments. The 12-symbol preamble ensured reliable automatic gain control, synchronization, and frequency offset compensation for multi-kilometer links with a fixed overhead of approximately $(12 + 4.25) \times 1.024$\,ms $\approx$ 16.6\,ms at SF7.

During operation, the transceiver remained in Sleep mode and transitioned through Standby to TX mode only when data transmission was required. Following each transmission, the radio immediately returned to Sleep mode. Average system current consumption outside transmission windows was measured at $<$65\,$\mu$A. The DIO0 pin was mapped to the TxDone interrupt, enabling interrupt-driven completion: the MCU entered sleep mode during transmission, the interrupt service routine set a completion flag upon TxDone assertion, and the main control loop subsequently cleared flags and returned the radio to Sleep mode. This approach eliminated polling overhead and minimized CPU active time.

The first-byte SPI read anomaly mitigation (single warm-up dummy read) is performed only once during initialization and not repeated in each transmission cycle, avoiding repetitive overhead on a per-packet basis.

\begin{table*}[h]
    \centering
    \resizebox{\textwidth}{!}{%
    \begin{tabular}{@{}l|l|l|l|l|l|l|l@{}}
    \toprule
    \textbf{Payload (B)} & \textbf{Driver} & $I_{\text{tx,avg}}$ (mA) & $I_{\text{idle}}$ (mA) & $I_{\text{cycle}}$ (mA) & Tx time (ms) & $E_{\text{tx}}$ (mJ) & $E_{\text{cycle}}$ (mJ) \\
    \midrule
    \rowcolor[HTML]{F3F3F3}
    \multirow{2}{*}{\textbf{1}} & Ours & \textbf{58.968 $\pm$ 0.027} & \textbf{0.065$\pm$ 0.004} & \textbf{1.768 $\pm$ 0.004} & \textbf{30.122 $\pm$ 0.013} & \textbf{5.862} & \textbf{5.994} \\
    
    & Standard & 59.866 $\pm$ 0.010 {\color[HTML]{D00000}(\textbf{+1.5\%})} & 1.960 $\pm$ 0.000 {\color[HTML]{D00000}(\textbf{+2938.8\%})} & 3.610 $\pm$ 0.000 {\color[HTML]{D00000}(\textbf{+104.2\%})} & 29.950 $\pm$ 0.025 {\color[HTML]{008000}(\textbf{-0.6\%})} & 5.917 {\color[HTML]{D00000}(\textbf{+0.9\%})} & 12.241 {\color[HTML]{D00000}(\textbf{+104.2\%})} \\
    
    \midrule
    \rowcolor[HTML]{F3F3F3}
    \multirow{2}{*}{\textbf{2}} & Ours & \textbf{59.148 $\pm$ 0.025} & \textbf{0.061 $\pm$ 0.003} & \textbf{2.050 $\pm$ 0.000} & \textbf{35.500 $\pm$ 0.012} & \textbf{6.929} & \textbf{6.984} \\

    & Standard & 59.870 $\pm$ 0.021 {\color[HTML]{D00000}(\textbf{+1.2\%})} & 1.952 $\pm$ 0.004 {\color[HTML]{D00000}(\textbf{+3079.2\%})} & 3.878 $\pm$ 0.007 {\color[HTML]{D00000}(\textbf{+89.2\%})} & 35.050 $\pm$ 0.042 {\color[HTML]{008000}(\textbf{-1.3\%})} & 6.925 {\color[HTML]{008000}(\textbf{-0.1\%})} & 13.218 {\color[HTML]{D00000}(\textbf{+89.2\%})} \\
    
    \midrule
    \rowcolor[HTML]{F3F3F3}
    \multirow{2}{*}{\textbf{16}} & Ours & \textbf{59.302 $\pm$ 0.026} & \textbf{0.060$\pm$ 0.004} & \textbf{3.210 $\pm$ 0.000} & \textbf{56.836 $\pm$ 0.011} & \textbf{11.123} & \textbf{11.158} \\

    & Standard & 59.938 $\pm$ 0.012 {\color[HTML]{D00000}(\textbf{+1.1\%})} & 1.956 $\pm$ 0.005 {\color[HTML]{D00000}(\textbf{+3160.0\%})} & 5.020 $\pm$ 0.006 {\color[HTML]{D00000}(\textbf{+56.4\%})} & 55.258 $\pm$ 0.075 {\color[HTML]{008000}(\textbf{-2.8\%})} & 10.930 {\color[HTML]{008000}(\textbf{-1.7\%})} & 17.451 {\color[HTML]{D00000}(\textbf{+56.4\%})} \\
    
    \bottomrule
    \end{tabular}%
    }
    \vspace{0.1cm}
    \caption{Comparison of our optimized low-power versus standard LoRa driver implementations per transmission across different payload sizes (N=10). Percentages indicate relative difference from the low-power implementation.}
    \vspace{-0.3cm}
    \label{tab:lora_driver_comparison}
\end{table*}

\clearpage
\section{Early Language Development and Phoneme Perception}\label{app:children}

Understanding how young children acquire language is a central question in developmental science. To study early language and cognitive development, developmental scientists rely largely on longitudinal naturalistic data, including speech, vocalizations, gestures, and social interactions between children and caregivers. These data are often collected continuously and passively in real-world environments, enabling the capture of rich and spontaneous behaviors that are difficult to observe in laboratory settings. However, such continuous in-the-wild data collection introduces significant design challenges, requiring the joint optimization of battery life, storage footprint, and on-device analysis capabilities for specialized applications, while maintaining a lightweight, portable, and minimally obtrusive form factor during daily activities.

\emph{We collaborated with developmental researchers} to understand real-world data collection needs and developed a lightweight, comfortable, and accessible head-worn device for capturing children's development data from an egocentric perspective. By jointly optimizing for power consumption, storage capacity, and thermal constraints, we enable continuous real-world data collection in naturalistic settings. This supports applications such as longitudinal observation of infant–caregiver interactions, phoneme perception in child-focused speech, and early intervention for delayed language acquisition. 

\begin{figure}[h]
    \centering
    \includegraphics[width=\linewidth]{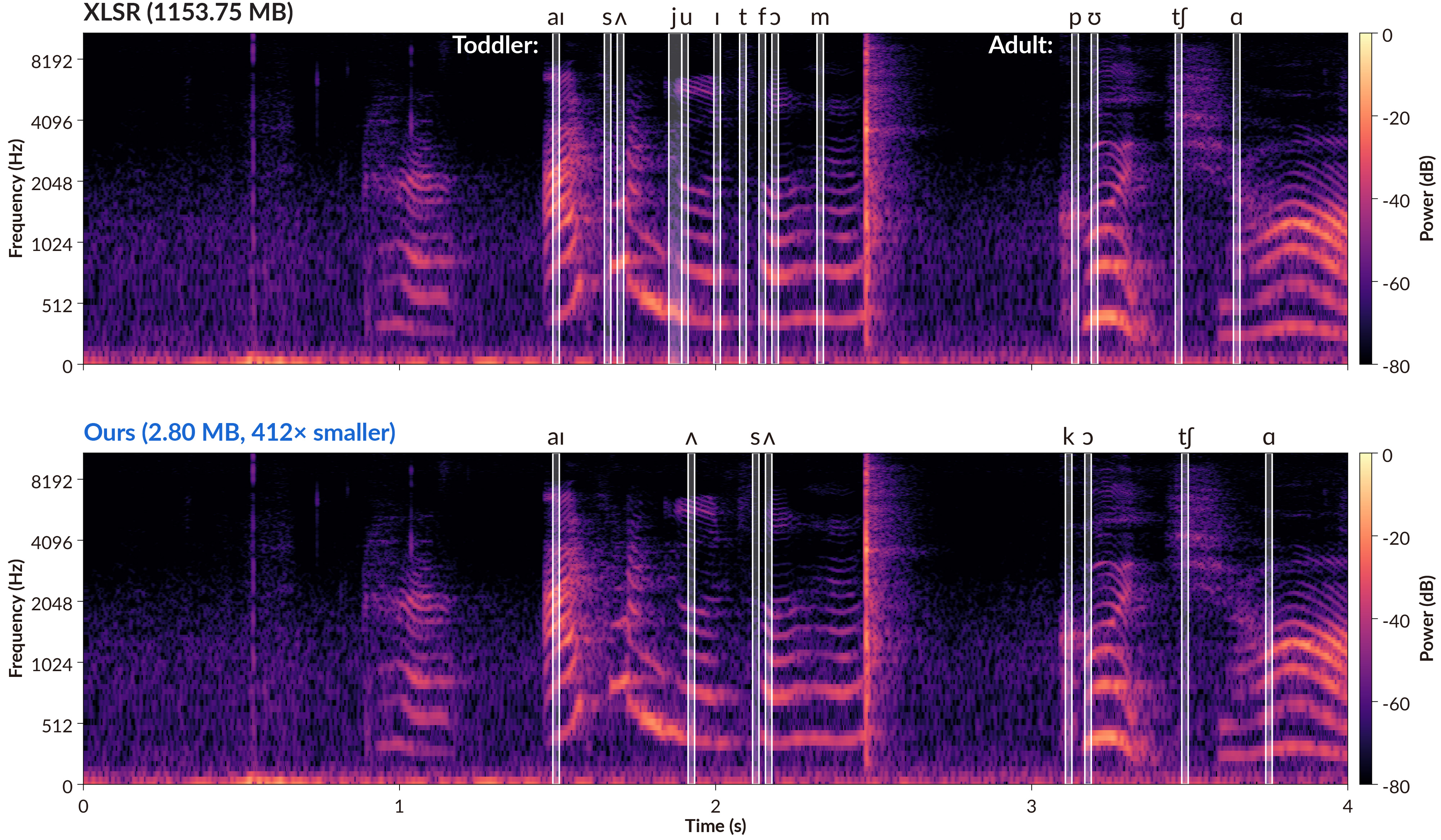}
     \caption{Log-mel spectrograms of a 4-second audio snippet of toddler and caregiver speech, overlaid with timestamped phonetic transcriptions from the full-size XLSR model and our compressed model. The toddler says \emph{``I can do it for him''}, pauses briefly, and the caregiver responds \emph{``good job''}.
     }
    \label{fig:infant_wearable_audio}
\end{figure}

As a preliminary demonstration, we present a 4-second audio snippet containing speech from a toddler and a caregiver, recorded with the wearable amid background noise from several children playing with toys. The audio snippet is de-identified, unpublished data provided by an external collaborator and collected under a university IRB-approved study.
The toddler says \emph{``I can do it for him''}, pauses briefly, and the caregiver responds \emph{``good job''}.
Fig.~\ref{fig:infant_wearable_audio} compares the phonetic transcriptions produced by the full-size XLSR model (1153.75\,MB) and our compressed model (2.80\,MB, $412\times$ smaller).
Both models correctly localize the two speech regions and the intervening pause.
On the caregiver's adult speech, the compressed model performs comparably to XLSR. Both transcribe ``job'' as \textipa{/tSA/}, and the compressed model's \textipa{/kO/} for ``good'' is arguably closer to the canonical \textipa{/gU/} than XLSR's \textipa{/pU/}, since \textipa{/k/} shares the velar place of articulation of \textipa{/g/}.
On the toddler's utterance, the gap widens considerably. XLSR recovers most of the sentence as \textipa{/aI s2 ju I t fO m/}, and its errors are attributable to child articulation: the onset of ``can'' becomes \textipa{/s/} rather than \textipa{/k/}, as child stop consonants are often weakly released, and ``do'' becomes \textipa{/ju/}, reflecting a palatalized production in which the alveolar stop approaches a palatal glide.
The compressed model, by contrast, detects only four phonemes (\textipa{/aI 2 s2/}). It identifies ``I'' correctly but omits most of the remaining segments and substitutes \textipa{/s2/} for ``for.''

The errors on the toddler's utterance share a common cause: the acoustics of a child's vocal tract differ from those of the adult speech on which both models were trained. Future work can therefore focus on fine-tuning with real child recordings collected by our deployed wearables.
Future improvements can focus on fine-tuning using real child recordings collected by our deployed wearables. Such adaptation could support applications such as real-time pronunciation feedback, in which caregivers receive alerts about mispronounced phonemes and can provide immediate correction, an important capability for early language learning.

\begin{figure}[h]
  \centering
  \includegraphics[width=0.5\linewidth]{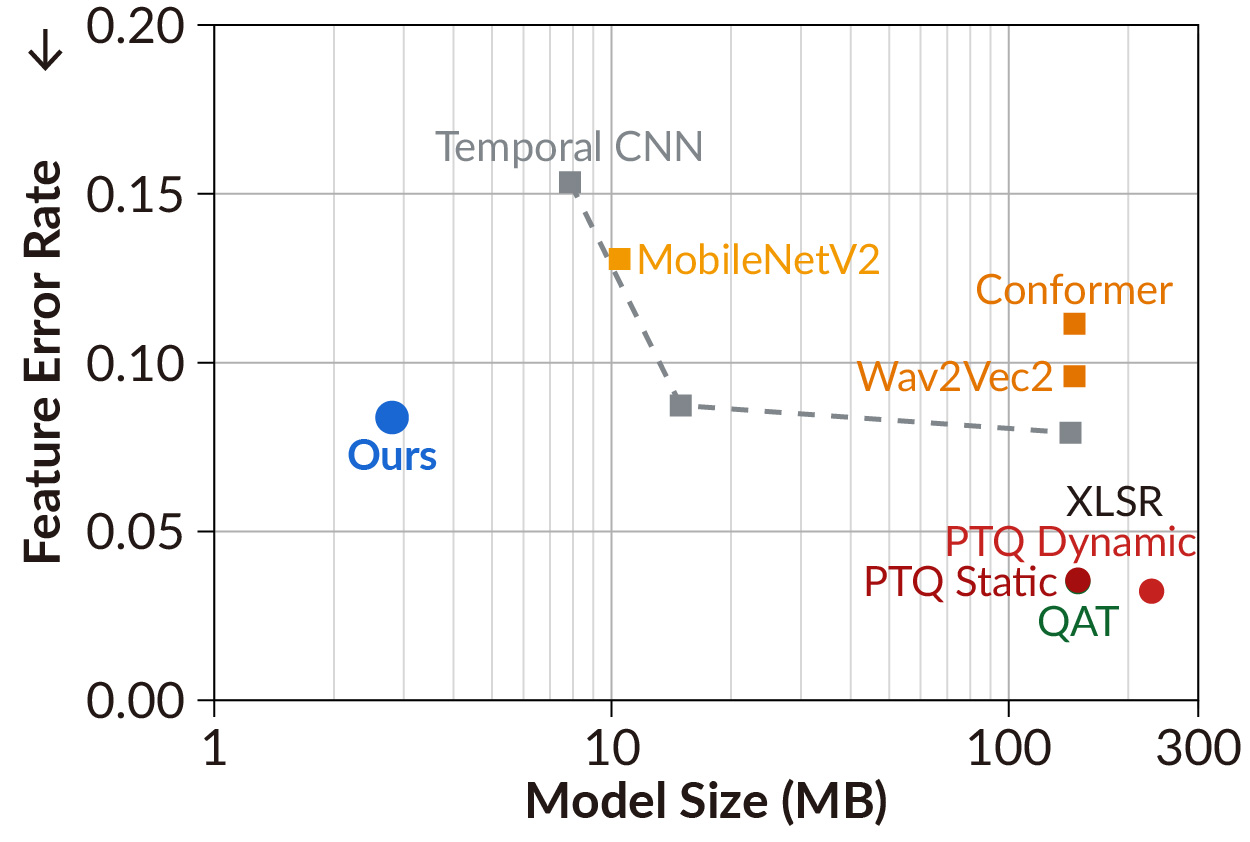}
  \caption{Comparison of our compressed models with standard compression baselines for speech-to-IPA transcription.} 
   \vspace{-1em}
  \label{fig:application_child_baselines}
\end{figure}

%%%%%%%%%%%%%%%%%%%%%%%%%%%%%%%%%%%%%%%%%%%%%%%%%%%%%%%%%%%%

% \clearpage
% \input{checklist.tex}

\end{document}

%% file: tables/compression_best_map_max78000.tex
\begin{wraptable}{r}{0.54\columnwidth}
    \centering
    \definecolor{customcellcolor}{HTML}{1967D2}
    \definecolor{highercolor}{HTML}{1967D2}
    \small
    \vspace{-1em}
    \caption{\textbf{Best On-Device Accuracy on MAX78000.} Values are mAP50-95 (YOLO11, COCO); \textcolor{highercolor}{\emph{higher}} is better. \redx\ indicate no successful deployment within $N_{\max}$ attempts. Three HIL agents surpass the human expert baseline (14.15\%).}
    \label{tab:compression_best_map_max78000}
    \begin{tabular}{@{}lcccccc@{}}
    \toprule
    \textbf{Model} & \multicolumn{2}{c}{\textbf{Sco}} & \multicolumn{2}{c}{\textbf{Doc}} & \multicolumn{2}{c}{\textbf{HIL}} \\
    \cmidrule(lr){2-3} \cmidrule(lr){4-5} \cmidrule(lr){6-7}
    \multicolumn{1}{r}{\textit{Thinking}} & \textit{L} & \textit{H} & \textit{L} & \textit{H} & \textit{L} & \textit{H} \\
    \midrule
    Claude Opus 4.7   & \redx & \redx  & \redx & \redx                         & \cellcolor{customcellcolor!15}15.23 & \cellcolor{customcellcolor!17}\textbf{16.93} \\
    Claude Sonnet 4.6 & \redx & \redx  & \redx & \redx                         & \redx                              & \redx   \\
    GPT-5.4           & \redx & \redx  & \redx & \cellcolor{customcellcolor!8}8.42 & \cellcolor{customcellcolor!9}9.12  & \cellcolor{customcellcolor!12}\textbf{12.07} \\
    GPT-5.4 mini      & \redx & \redx  & \redx & \redx                         & \redx                              & \redx   \\
    Gemini 3.1 Pro    & \redx & \redx  & \redx & \redx                         & \cellcolor{customcellcolor!10}10.43 & \cellcolor{customcellcolor!18}\textbf{18.03} \\
    Gemini 3 Flash    & \redx & \redx  & \redx & \redx                         & \redx                              & \redx   \\
    \bottomrule
    \end{tabular}
    % \vspace{-1em}
\end{wraptable}

%% file: tables/compression_best_fer_stm32n6.tex
\begin{wraptable}{r}{0.65\columnwidth}
    \centering
    \definecolor{customcellcolor}{HTML}{A8DAB5}
    \definecolor{lowercolor}{HTML}{188038}
    \small
    \vspace{-1em}
    \caption{\textbf{Best On-Device Error on STM32N6.} Values are Feature Error Rate (\%; Wav2Vec2, speech-to-IPA); \textcolor{lowercolor}{\emph{lower}} is better.}
    \label{tab:compression_best_fer_stm32n6}
    \begin{tabular}{@{}lcccccc@{}}
    \toprule
    \textbf{Model} & \multicolumn{2}{c}{\textbf{Sco}} & \multicolumn{2}{c}{\textbf{Doc}} & \multicolumn{2}{c}{\textbf{HIL}} \\
    \cmidrule(lr){2-3} \cmidrule(lr){4-5} \cmidrule(lr){6-7}
    \multicolumn{1}{r}{\textit{Thinking}} & \textit{L} & \textit{H} & \textit{L} & \textit{H} & \textit{L} & \textit{H} \\
    \midrule % round((cell value - 25) * 2):
    Claude Opus 4.7   & \cellcolor{customcellcolor!27}38.69 & \redx & \cellcolor{customcellcolor!30}39.90 & \redx & \cellcolor{customcellcolor!14}32.14 & \cellcolor{customcellcolor!8}\textbf{29.12} \\
    Claude Sonnet 4.6 & \cellcolor{customcellcolor!27}38.41 & \redx & \cellcolor{customcellcolor!35}42.31 & \redx & \cellcolor{customcellcolor!10}\textbf{29.78} & \redx \\
    GPT-5.4           & \redx & \cellcolor{customcellcolor!24}36.93 & \cellcolor{customcellcolor!34}41.81 & \cellcolor{customcellcolor!24}37.01 & \cellcolor{customcellcolor!29}39.47 & \cellcolor{customcellcolor!21}\textbf{35.73} \\
    GPT-5.4 mini      & \redx & \redx & \redx & \redx & \redx & \cellcolor{customcellcolor!38}\textbf{44.29} \\
    Gemini 3.1 Pro    & \redx & \cellcolor{customcellcolor!17}33.57 & \redx & \cellcolor{customcellcolor!28}39.30 & \redx & \cellcolor{customcellcolor!14}\textbf{32.19} \\
    Gemini 3 Flash    & \redx & \redx & \redx & \redx & \redx & \redx \\
    \bottomrule
    \end{tabular}
    \vspace{-1em}
\end{wraptable}

%% file: tables/success_by_task.tex
\begin{wraptable}{r}{0.56\columnwidth}
    \centering
    \definecolor{customcellcolor}{HTML}{AECBFA}
    \definecolor{highercolor}{HTML}{1967D2}
    \small
    \vspace{-1em}
    \caption{\textbf{Deployment Success Rate by Task and Feedback Configuration.} Values are mean successful iterations (\%) across models ($N=6$); \textcolor{highercolor}{\emph{higher}} is better.}
    \label{tab:success_by_task}
    \begin{tabular}{lccc}
    \toprule
    \textbf{Task} & \textbf{Sco} & \textbf{Doc} & \textbf{HIL} \\
    \midrule
    \multicolumn{4}{l}{\textit{Power Minimization (MAX78000)}} \\
    \quad $I_{\text{peak}} \leq 32$\,mA & \cellcolor{customcellcolor!16}$16.7$ & \cellcolor{customcellcolor!38}$38.3$ & \cellcolor{customcellcolor!47}$\mathbf{46.7}$ \\
    \quad Energy        & \cellcolor{customcellcolor!25}$25.0$ & \cellcolor{customcellcolor!38}$38.3$ & \cellcolor{customcellcolor!68}$\mathbf{68.3}$ \\
    \midrule
    \multicolumn{4}{l}{\textit{Thermal Management (ESP32-S3R8)}} \\
    \quad Room temp ($T_{\text{peak}} \leq 37\,^\circ$C & \cellcolor{customcellcolor!13}$13.3$ & \cellcolor{customcellcolor!10}$10.0$ & \cellcolor{customcellcolor!27}$\mathbf{26.7}$ \\
    \quad Contact-heated ($T_{\text{peak}} \leq 51\,^\circ$C) & \cellcolor{customcellcolor!10}$10.0$ & \cellcolor{customcellcolor!17}$16.7$ & \cellcolor{customcellcolor!33}$\mathbf{33.3}$ \\
    \bottomrule
    \end{tabular}
    \vspace{-1em}
\end{wraptable}